# Electrical Degradation in Dielectric and Piezoelectric Oxides: Review of Defect Chemistry and Associated Characterization Techniques

Pedram Yousefian,[a,*] Betul Akkopru-Akgun,[a] Clive A. Randall,[a] and Susan Trolier-McKinstry [a]

[a] Materials Research Institute, Department of Materials Science and Engineering, Center for Dielectrics and Piezoelectrics, The Pennsylvania State University, University Park, PA, 16802, USA

## 1 Abstract

The properties of dielectric and piezoelectric oxides are determined by their processing history, crystal structure, chemical composition, microstructure, dopants (or defect) distribution, and defect kinetics. These materials are essential in a diverse range of applications including aerospace, medical, military, transportation, power engineering, and communication, where they are used as ceramic discs, thick and thin films, multilayer devices, etc. Significant advances in understanding the materials, processing, properties, and reliability of these materials have led to their widespread use in consumer electronics, military, and aerospace applications. This review delves into electrical degradation in dielectrics and piezoelectrics, focusing on defect chemistry and key characterization techniques. It also provides a detailed discussion of various spectroscopic, microscopic, and electronic characterization techniques essential for analyzing defects and degradation mechanisms.

## 2 Introduction

Perovskite ceramics, known for their ferroelectric, dielectric, piezoelectric, and pyroelectric properties, are utilized in various applications, including multilayer ceramic capacitors (MLCC) [1–5], piezoelectric Micro-Electro-Mechanical Systems (MEMS) [6–12], positive temperature coefficient resistors (PTCR) [13–19], sensors [20–26], actuators [24,27–32], solid oxide fuel cells [33,34], and electrocaloric cooling [35–37]. As these devices continue to miniaturize, the thickness of the insulating oxide layers decreases, resulting in increased electrical field stress under an applied bias. This escalation can lead to device failure through mechanisms such as thermal breakdown from excessive current, dielectric breakdown, or electrical degradation characterized by increasing leakage current under a DC bias and thermal stress. Electrical degradation in dielectrics and piezoelectrics poses a significant challenge, critically affecting the longevity and reliability of electronic devices. Central to this issue is defect chemistry and migration of mobile point defects. This scientific domain is key to understanding and addressing the electrical behavior of dielectrics and piezoelectrics. Such defects, encompassing everything from atomic vacancies, impurity atoms, to multivalent cations states can lead to reduced performance or failure over time. The performance of electroceramics relies on their responses to electric fields, and thus, comprehending and countering electrical degradation is a complex and important task. It requires a deep understanding of the inherent defect chemistry and their dynamics as well as proficiency in characterization and modeling techniques to effectively identify, analyze, and predict the behavior of these defects.

One typical manifestation of dielectric degradation on prolonged exposure of a dielectric to a DC field is illustrated in Figure 1 [38]. The current response over time reveals three distinct regimes. Following the charging of the capacitor, there is a transitory regime marked by an initial decay in current as the material adjusts to the electric field. This provides a relaxation current that can be mathematically expressed by the Curie-von Schweidler law, indicative of a power-law decay:

$$I_R(t) = I_0 t^{-n} \tag{1}$$

where $I_0$, t, and n are initial current, time, and a constant, respectively. This regime includes a combination of space charge polarization, also known as Maxwell-Wagner polarization, which arises due to inhomogeneous electrical conductivity characterized by a broad spectrum of relaxation times; charge trapping and space charge formation near the ionically-blocking metal electrodes (permeable to the electronic carriers) which reduce the electrical field and injected current; and a superposition of Debye-type relaxations that encompasses a range of relaxation times [39–42].

Following this is the steady-state or leakage current region, a regime where a constant leakage current persists in the material under the electric field for a substantial length of time. This region is crucial for identifying the dominant conduction mechanism, a topic thoroughly examined in various review papers and books [3,43–46]. In many capacitive ceramics, the voltage drop occurs across grains, grain boundaries and interfaces, with the combined resistance of these elements dictating the steady-state current. Particularly in some ceramics, double Schottky barriers at grain boundaries and interfaces control this steady-state leakage current [47].

The final phase toward the end of life is the electrical degradation region, in which prolonged electric field exposure leads to an increased leakage current, ultimately contributing to material failure. This electrical degradation is a significant factor in the failure of dielectric and piezoelectric materials, underscoring the necessity to understand and mitigate the mechanics behind this phenomenon.

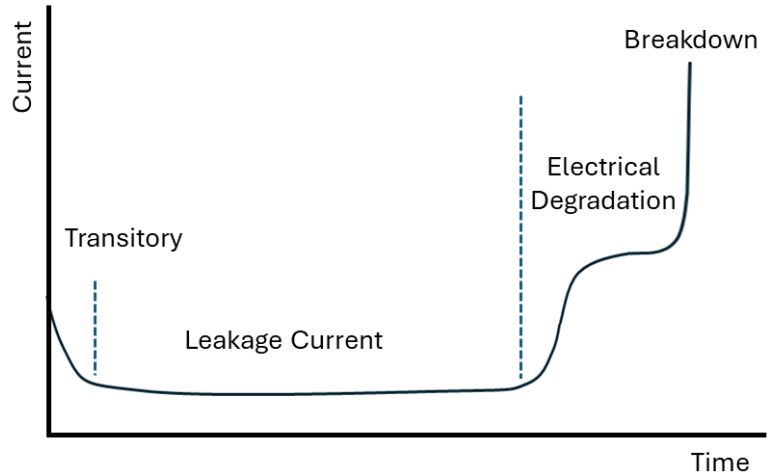

Figure 1. Schematic of the current response of a dielectric material subjected to an applied DC electric field at an elevated temperature demonstrates three separate regimes: the transitory regime, the leakage current regime, and the electrical degradation regime [38].

This review is dedicated to examining the critical regime of electrical degradation in dielectrics and piezoelectrics, with a focus on the underlying defect chemistry and important characterization techniques used to probe it. Designed to be a guide for both researchers and practitioners, this overview bridges the gap between the defect chemistry and the tangible challenges of mitigating electrical degradation in these materials. The review delves into the roles of various defect types in electrical degradation, examining their thermodynamics and kinetics. This is critical to the processes of defect formation, migration, and interaction within these materials. Moreover, the review dedicates a substantial section to highlighting a wide array of characterization techniques used to characterize defect chemistry and degradation mechanisms. From state-of-the-art spectroscopic and microscopic techniques to the broader perspectives offered by electronic techniques, each method contributes significantly to the collective understanding of defects and their impacts. The intent of the paper is to provide a comprehensive view of the current state of knowledge, while also identifying promising avenues for future research. Ultimately, this review aspires to empower those in the field with the knowledge and tools to combat the challenges posed by electrical degradation, thus advancing the reliability and efficacy of dielectrics and piezoelectrics in a wide array of applications.

# 3 The Role of Defect Chemistry on Electrical Degradation

The role of defect chemistry in electrical degradation is critical to the dielectrics and piezoelectrics that are integral to modern electronic devices. Despite significant research aimed at enhancing the quality of these devices through defect chemistry and interface engineering, the complete elimination of defects remains elusive. These microscopic defects, such as vacancies, interstitials, or impurity atoms, disproportionately affect the electrical properties and stability of materials. A comprehensive understanding of these defects is essential in combating electrical degradation, which can adversely affect the performance and longevity of electronic components. Delving into the intricacies of defect chemistry not only sheds light on the complex electrical degradation mechanisms but also paves the way for improving material properties through defect engineering. In this section, an overview of the defect chemistry in key bulk and thin film dielectric and piezoelectric materials is presented. This overview includes an exploration of various strategies for mitigating electrical degradation in each material system. The exploration of these imperfections links fundamental scientific research and the practical demands of technological progress, underscoring its importance in both academic and industrial spheres.

## 3.1 Barium Titanate, Strontium Titanate, and Barium Strontium Titanate

### 3.1.1 Defect Chemistry

Perovskite titanates, such as barium titanate ($BaTiO_3$ or BTO) and strontium titanate ($SrTiO_3$ or STO), have exceptional ferroelectric and dielectric properties which are linked to their crystal structures [3,22,30,35,48,49]. However, commercial implementation of these materials requires an understanding of their electrical degradation, which is a strong function of the defect chemistry. Exploration of the defect chemistry of both undoped and doped $(Ba,Sr)TiO_3$, has been undertaken by both experimental [50–87] and theoretical [88–99] investigations. Here, the crystal structures and defect chemistry of several perovskite titanates are reviewed, prior to a more in-depth review of characterization methodologies needed to elucidate the underlying degradation mechanisms.

The perovskite crystal structure of $(Ba,Sr)TiO_3$ features a twelve-fold coordinated A-site occupied by the $Ba^{+2}/Sr^{+2}$ ion, an octahedrally coordinated B-site occupied by $Ti^{+4}$, and the anion site occupied by $O^{-2}$ [100]. The dense atomic packing in the perovskite structure favors the formation of Schottky defects over Frenkel defects. The Schottky reaction, balancing charge, mass, and sites, can be written as:

$$MTiO_3 \leftrightarrow V_M'' + V_{Ti}'''' + 3V_O^{\bullet\bullet} \tag{2}$$

where M denotes Ba or Sr. The oxygen vacancies ($V_O^{\bullet\bullet}$) are doubly charged with respect to the lattice; this has been found to be energetically preferred over singly ionized oxygen vacancies [101]. In these ternary oxides, the chemistry can be considered as quasi-binary mixtures of two sub-oxide units (MO)(TiO$_2$), accommodating both full and partial Schottky defect reactions. While full Schottky defects maintain stoichiometry, partial Schottky reactions generate electronic carriers, specifically conducting electrons and holes. These reactions are non-stoichiometric, altering the (Ba,Sr)/Ti ratio and oxygen-deficiency within the single-phase field [48,50,52,53,77,85,102–105].

In perovskite titanates like STO, defect behaviors vary with temperature. Below 750 K, oxygen reduction slows, stabilizing strontium and oxygen vacancy concentrations in the lattice. Above 770 K, oxygen reduction reactions intensify, affecting vacancy mobility, while full or partial Schottky defect concentrations tend to be frozen below approximately 1470 K due to limited cation vacancy diffusion [17,48]. These temperature-dependent dynamics of oxygen and Schottky defects are crucial for understanding and manipulating the properties of perovskite titanates. In addition to temperature, the defect chemistry is also influenced by the oxygen partial pressure ($P_{O_2}$), dictating the prevalence of specific defect reactions. Brouwer diagrams provide a graphical representation of defect concentrations across the spectrum from oxidation to reduction conditions [84,106]. Figure 2 presents the Brouwer diagrams of undoped BTO and acceptor-doped STO, showcasing striking similarities in electron and hole concentration changes as a function of $P_{O_2}$.

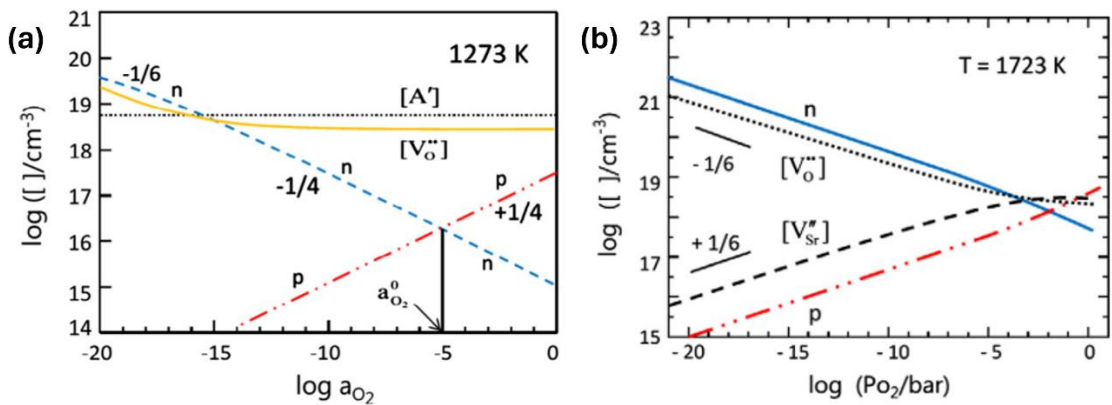

Figure 2. Brouwer schematic diagrams of full Schottky defect reaction for undoped (a) BaTiO$_3$ [74], (b) SrTiO$_3$ [54]. Reproduced with permission from ref. [48].

Point defects in BTO and STO are regarded similarly, with minimal adjustments in the models used to describe their defects and mechanisms [48,54,63,107]. However, differences in the solubility of (Ba,Sr)/Ti in their Brouwer diagrams implies that there are variations in metal vacancies and their concentrations, leading to distinct defect chemistries at different oxygen partial pressures. This distinction has been supported by first-principles Density Functional Theory (DFT) calculations [98,108,109]. The challenge arises from the inherent difficulty in directly observing the effects of metal vacancies in these materials [51,110,111]. In STO, strontium and oxygen vacancies are predominant due to their low formation energy, impacting carrier mobility and conductivity [52,112–116]. On the other hand, BTO has been the focus of extensive research regarding oxygen vacancies, Schottky defects, and metal vacancies induced by donor and acceptor dopants [53,55,67,71,74,84,103,104,110,117–124]. Investigations into solubility regimes have led to a better understanding of defect formation energies for various Schottky reactions in BTO [77,85,125]. These studies indicate that under low oxygen partial pressure conditions, defect formation energies increase, decreasing the metal vacancy concentration. A key finding from these studies is that at extremely low oxygen partial pressures and high temperatures, stoichiometric BTO undergoes reduction, leading to the formation of oxygen vacancies and the conversion of $Ti^{4+}$ to $Ti^{3+}$. At intermediate oxygen partial pressures and elevated temperatures, stoichiometric BTO shows a tendency towards the Schottky defect reaction, while non-stoichiometric BTO tends towards the partial Schottky defect reaction [85]. High-temperature DC-conductivity measurements have also revealed that metal vacancies in BTO can compensate for donor impurities, particularly during the reoxidation of intragranular interfaces. This compensation mechanism is more favorable at grain boundaries than within the grains themselves. This insight, supported by thermogravimetric studies [126], indicates that metal vacancies at grain boundaries create resistive back-to-back Schottky-type barriers, affecting the overall electrical properties of BTO [17].

### 3.1.2    Electrical Degradation and Mitigating Strategies

In perovskite titanates, the kinetics of electrical degradation depend on the oxygen partial pressure, the quality of interfaces, electrode density, grain size, and the concentration of defects and impurities in the materials as prepared [127–131]. Electrical degradation can occur within the bulk dielectric as well as at the metal-dielectric interfaces. In the bulk, degradation is characterized by an increased total conductivity, due to the accumulation of defects, which results in regions with either high or low resistance. Key mechanisms that drive these changes include charge injection, ionic de-mixing, or the creation of conductive filaments [1,61–63,99,132–137].

Waser et al. [61–63] demonstrated that in STO, the electromigration of oxygen vacancies between cathode and anode leads to degradation; the degradation rate depended on dopant concentration, temperature, and the applied DC electric field [47]. This occurs as oxygen vacancies de-mix due to the blocking effect of the electrodes. A p-n junction is formed, decreasing the oxygen vacancy concentration in the anodic region and increasing it in the cathodic region. The dielectric resistance is degraded as this p-n junction is biased in the forward direction. The degradation process initiates as oxygen vacancies move across grains, leading to their accumulation at the grain boundaries, which changes the field distribution and increases the tunnelling current [47]. Moreover, Maier et al. [138–142] and Randall et al. [83,143,144] investigated local conductivity and defects in degraded Fe-doped and undoped STO using *in-situ* impedance spectroscopy, correlating activation energies with an acceptor-type trap model. In Fe-doped STO, the oxygen vacancies migrate from the anode to the cathode. Local charge compensation induced increased concentrations of holes and $Fe^{4+}$, favoring p-type conductivity near the anode, triggering an electrocoloration change [145,146]. Conversely, oxygen vacancies moving towards the cathode introduce more donors, resulting in higher electron concentrations.

Most models for electrical degradation overlook the potential impact of surface states and interphase states, assuming uniform defect concentrations throughout the material. However, this approach does not capture variations in individual defect energies at surfaces and grain boundaries, which lead to the formation of interfacial space charge [147–153]. Furthermore, not all ions exhibit the same mobility, and the cations are largely immobile in most perovskite oxides near ambient temperatures. Over time, this disparity can, in principle, lead to chemical inhomogeneity within the material under applied field. This phenomenon, known as kinetic unmixing, can theoretically push the material out of its stability range, leading to what is termed a kinetic decomposition. Kinetic unmixing in oxides has been reported to be driven by varying oxygen partial pressure [154], or an applied voltage [96,155–159]. Ader et al. [96] determined that both undoped and acceptor doped BTO are affected by kinetic unmixing, leading to concentration gradients among cations, oxygen ions, electrons, and holes. Their simulation results indicated potential kinetic decomposition at the cathode or anode, dependent on the initial cation composition; the lifetimes varied based on the initial cation composition and voltage parameters. The lengthy simulation timescales explain why kinetic decomposition is not observed *experimentally* in BTO. In BTO at room temperature, the high activation energies of cation diffusion (around 5 eV) result in extremely low diffusion coefficients. Thus, lifetime limits imposed by decomposition become significant only at temperatures exceeding ~1273 K. Figure 3 illustrates the non-uniform oxide boundary velocities in BTO at the cathode and anode; it can be seen that steady state is not attained even after ~ $5\times10^{14}$ seconds (~15 million years) [96].

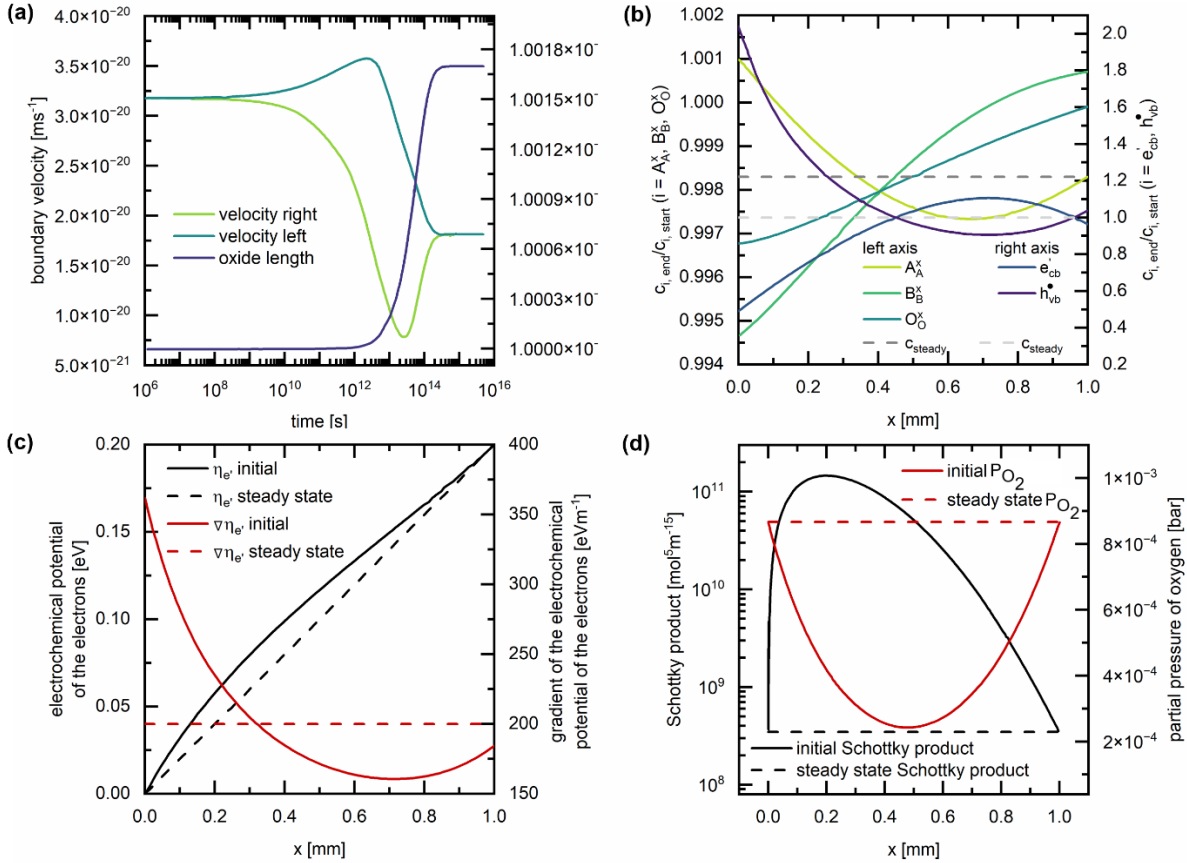

Figure 3. Simulation results for undoped BTO with coordinates relative to the anode (left) and cathode (right). (a) boundary velocities and oxide length changes over time, (b) concentration profiles normalized to initial values in the steady state, (c) electron electrochemical potential and gradient profiles, and (d) comparative Schottky product and oxygen partial pressure profiles. Redrawn from data in ref. [96].

Unlike bulk degradation, interface degradation usually involves a decrease in the Schottky barrier, which is often a result of charge injection. The accumulation of charged ionic species at the cathode metal/ceramic interface reduces both the Schottky barrier's height and the depletion region's width. These alterations accelerate electron conduction into the metal, resulting in a higher current density. Unlike in the bulk, charge compensation at interfaces is not limited. In ionic materials, like STO and BTO, defects preferentially accumulate at interfaces due to lower formation energies, leading to the formation of a space charge layer even prior to degradation. During sintering, acceptor dopants can segregate at the grain boundaries, resulting in space charge potentials at interfaces that influence grain growth and the conductivity of polycrystalline materials [148,152,153,160,161]. Impedance spectroscopy has revealed that grain boundaries form double Schottky barriers with space charge depletion layers, impeding the movement of oxygen vacancies across the boundaries. The role of grain boundaries as kinetic barriers diminishes with the decreasing number of grain boundaries in thin active layers, which may change the conduction mechanism from bulk dominated toward electrode/ceramic-interface dominated [162]. Grain microstructure and grain boundary segregation also influence the overall properties, with bulk defect chemistry dominating in larger grains (>5 μm) and grain boundaries dominating conductivity in smaller grained ceramics [124,163,164].

Interface engineering techniques, including grain boundary engineering and the use of interface layers, can be employed to control the behavior of oxygen vacancies and mitigate their impact on degradation [3,165–167]. In MLCCs with thin active layers, precise control of grain size and dopants distribution gains heightened significance, as any imperfections becomes critical as weak spots and can significantly impact overall performance [168]. Maintaining at least 5-7 grain boundaries perpendicular to the path of oxygen vacancies electromigration is necessary to mitigate electrical degradation [169]. These findings highlight the importance of understanding microstructure-electrical property interactions to effectively address degradation in dielectric materials.

Electrical degradation in $(Ba,Sr)TiO_3$ thin films, while similar to bulk materials, exhibits a stronger dependence on applied voltage and varied dopant concentration effects compared to bulk materials [112,170–178]. Some films show evidence for electromigration of oxygen vacancies [174,176,178]. In some films, the cathode electrode material significantly influences thin film degradation [170,179]. Comparisons with STO ceramics and single crystals reveal additional mechanisms in thin film degradation beyond the de-mixing process observed in bulk materials. In other films, the oxidation and reduction models are less applicable, as there is minimal influence from the ambient atmosphere and degradation is not exclusively caused by electronic charge carriers. This indicates that in some cases, thin film degradation depends heavily on the microstructure and interfaces [170,176,178].

The internal electrode interface, which is critical in controlling electron injection at the cathode, has a significant impact on the overall performance of BTO-based MLCCs [180]. The two primary types of BTO-based MLCCs are precious metal electrode (PME) and base metal electrode (BME) MLCCs., Ag, Pt, Pd, and/or Ag-Pd alloys are utilized as internal electrodes in PME MLCCs and can be sintered in ambient air. On the other hand, BME MLCCs employ Ni or Cu electrodes for cost considerations and are sintered under low oxygen partial pressures to prevent oxidation of the metal electrodes (Ni + $\frac{1}{2}O_2 \rightarrow$ NiO (e.g., $P_{O_2} < 10^{-10} MPa\ at\ 1400\ K$) [181–183]. With these electrodes, sintering reduces the BTO. Therefore, a subsequent re-oxidation annealing step at lower temperatures is performed to minimize the oxygen vacancy concentration while maintaining metal electrode integrity [81,82,180,184].

To address the electrical degradation resulting from oxygen vacancy electromigration, several strategies can be employed. Material engineering approaches encompass optimizing sintering conditions, controlling oxygen partial pressures, and doping. These approaches limit the formation and mobility of oxygen vacancies, hence minimizing degradation phenomena [3,76,81,82,185,186]. Perovskites like (Ba,Sr)TiO$_3$ are often modified with multiple dopants; the role of the dopant depends on its ionic size and charge [3,48,187]. Sakabe et al. [188] investigated the effect of rare-earth (RE) elements (Dy, Sm, La, and Yb) in BTO-based MLCCs using TEM and electron probe microanalysis (EPMA) which revealed La's tendency to occupy Ba sites, whereas Dy was present on both Ba and Ti sites. Notably, Dy doping significantly improved MLCC reliability, in contrast to La doping. Moriwake et al. [189] conducted a detailed analysis of RE elements (La, Dy, Ho) in BTO using first-principles calculations and thermodynamics and reported that La predominantly occupied Ba sites for a wide range of processing conditions. In contrast, Dy and Ho (which significantly improve the reliability of MLCC dielectric materials) occupied both Ba and Ti sites, in agreement with experimental observations. Sharma et al. [97] conducted a comprehensive first-principles investigation on various dopants in BTO. They found that dopants with an ionic radius approximately 30% larger than Ti and a nominal oxidation state of 2 or less, such as Co, Cu, Ni, and Zn at the Ti site, promote oxygen vacancy formation. Conversely, elements like Mo, W, Tc, V, Ta, Nb, and Re suppress oxygen vacancy formation [97]. While Ti site doping strongly influences oxygen vacancy formation, Ba site doping shows relatively limited control over oxygen vacancy concentration. Most A-site dopants act as oxygen vacancy suppressors, except for the Group IA dopants (K, Rb, and Cs), which promote oxygen vacancy formation due to the lower cation valence [97]. This behavior aligns with the electrical nature of dopants; p-type (n-type) dopants generally favor (hinder) oxygen vacancy formation. On the anion site, nitrogen doping in (Ba,Sr)TiO$_3$ films enhanced the permittivity, reduced dielectric loss, and improved electrical performance [178,190–195].

### 3.2 Lead titanate (PbTiO$_3$) and Lead zirconate titanate (PZT)

#### *3.2.1 Energy Band Characteristics:*

As a mixed electronic and ionic conductor, PZT has a band gap of 3.4 eV [196], with multiple in-gap states introduced by defects [45,46,196–201]. The valence band consists of O *2p* and Pb *6s* states, while the conduction band is made up of the unfilled d states of Ti$^{4+}$ [202]. The presence of lone electron pairs (6s electrons) allows Pb$^{2+}$ sites to act as shallow acceptor levels for holes. Early deep level transient spectroscopy (DLTS) studies reported by Smyth et al. confirmed that the Pb$^{2+}$/Pb$^{3+}$ level is located approximately 0.26-0.3 eV above the valence band edge [203]. Lead vacancies, created due to evaporation of PbO(g) or addition of donor ions, also introduce singly and doubly ionized lead vacancies that lie 0.9 and 1.4 eV above the valence band, respectively [204,205]. In the same way, introduction of acceptor ions creates additional impurity levels in the PZT energy band diagram. Redox active Mn ions produce two distinct impurity levels; (1) Mn$^{2+}$/Mn$^{3+}$ and (2) Mn$^{3+}$/Mn$^{4+}$ with ionization energies of 1.9 and 1.2 eV, respectively [206]. The main electron trapping centers in PZT are Ti$^{4+}$/Ti$^{3+}$ [203,207,208], which produce a level about 1 eV below the conduction band, and oxygen vacancies, which are shallow traps located about 0.2 eV below the conduction band [196]. These defect levels are shown in Figure 4.a.

In lead titanate, the valence band maximum (VBM) is located at the X point of the Brillouin zone, exhibiting a hybridized character that arises from lead s-orbitals and oxygen p-orbitals. Conversely, the conduction band minimum (CBM) is positioned at the Z point, characterized predominantly by the d$_{xy}$ orbital of titanium [209,210]. As the Zr/Ti ratio increases, there is a transition in the character of the CBM from being dominated by Ti/Zr d$_{xy}$ orbitals to Pb p-orbitals. The change in conduction-band character with increasing Zr/Ti content leads to (1) an increase in the trap binding energy at the Pb$^{2+/3+}$ center; the paramagnetic Pb$^{3+}$ hole center becomes deeper (Figure 4.b), and (2) an 0.2 eV increase in band gap in the Pb(Ti,Zr)O$_3$ in contrast to the significant 2 eV elevation in band gap of Ba(Ti$_{1-x}$Zr$_x$)O$_3$ with x [203].

Apart from the influence of the Ti/Zr ratio, it was observed that biaxial tensile strain significantly alters the energy band properties [210,211]. Insights from first-principles calculations show that tensile strain leads changes PbTiO$_3$ from tetragonal to monoclinic, and eventually to orthorhombic. This strain-induced phase change directly affects the electronic band structure, particularly impacting the conduction band and inducing spin splittings due to the spin-orbit interactions (Figure 4.c). Notably, in the monoclinic phase, spin splitting is intensified as a result of polarization rotation, with the maximum splitting occurring at a critical strain where the energy bands anticross (Figure 4.d) [212].

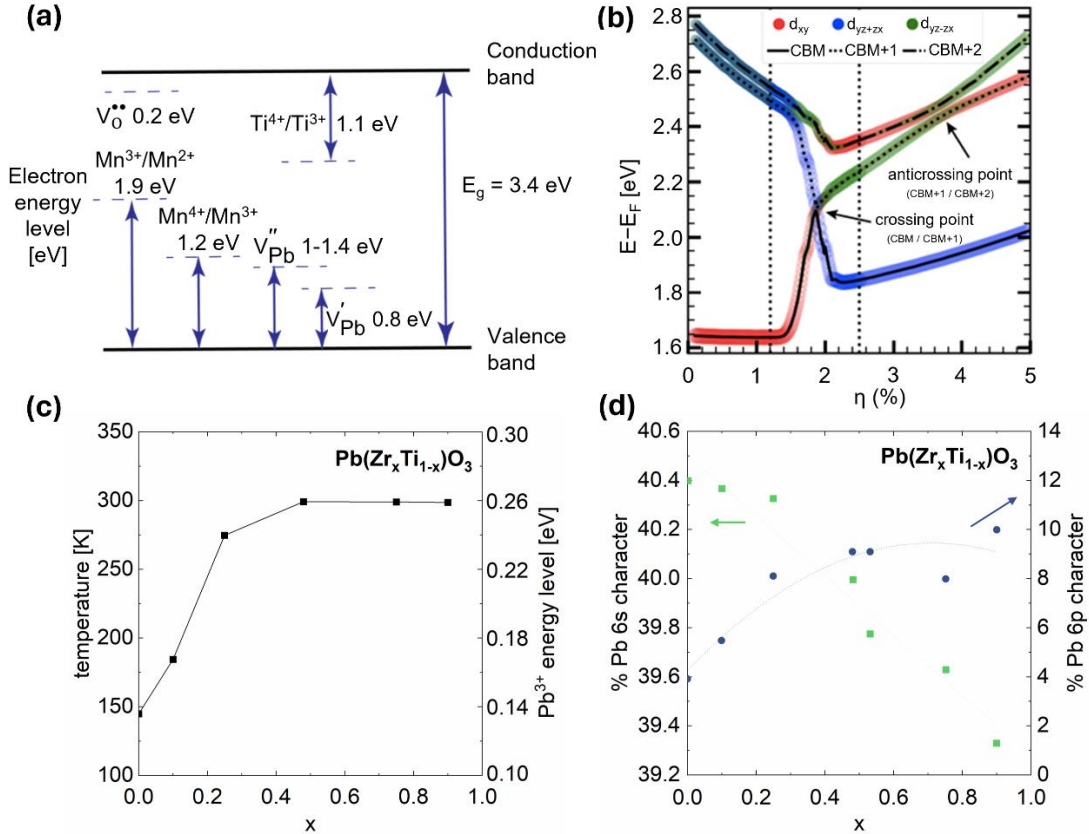

Figure 4. (a) Schematic representation of defect states in energy band diagram of PZT (reproduced with permission from ref. [6]), (b) Evolution of the energy of the three lowest conduction bands as a function of the strain η. The labels "CBM," "CBM + 1," and "CBM + 2" refer respectively to the first, second, and third conduction bands, ordered by increasing energy (reproduced with permission from ref. [211]), (c) Temperature stability and deduced energy level of the $Pb^{3+}$ center vs. Zr content. (d) Percentage Pb 6s and 6p character in the $Pb^{3+}$ center for various Zr contents (redrawn from data in ref. [212].

### 3.2.2 Defect Chemistry of PZT:

The compact arrangement of the perovskite lattice inhibits the formation of interstitial ionic defects, aside from hydrogen. Consequently, the primary defects in phase pure PZT encompass oxygen vacancies, lead vacancies, as well as electronic carriers such as electrons and holes [212–217]. Given that vacancies are the sole plausible lattice imperfections, the predominant form of intrinsic ionic disorder is likely to be Schottky-type defects:

$$nil \leftrightarrow V''_{Pb} + 3V^{\bullet\bullet}_O + V''''_{Ti,Zr} \qquad (3)$$

where nil denotes the defect-free crystal. Vacancies must arise in a stoichiometric proportion to maintain the charge neutrality. The defect concentrations are governed by a mass-action constant:

$$K_s(T) = [V''_{Pb}][V^{\bullet\bullet}_O]^3[V''''_{Ti,Zr}] = K'_s \cdot \exp(-\Delta H_s)/kT \qquad (4)$$

where $K_s(T)$ is the temperature-dependent mass-action constant, determined by the standard enthalpy ($\Delta H_S$) of the Schottky disorder reaction [217]. Theoretical predictions suggest a high value of the Schottky formation enthalpy [123]. Consequently, the lattice defects are primarily attributed to the presence of naturally occurring impurities, rather than to intrinsic disorder, except for temperatures near the melting point.

The defect chemistry of PZT differs from BaTiO$_3$ or SrTiO$_3$ due to the high levels of volatility of the PbO, which makes it difficult to control the stoichiometry precisely. The structure can tolerate a substantial loss of lead, up to 2 mol% [218,219]. When PbO evaporates, charge compensation is achieved either by creation of oxygen vacancies or by change in the concentrations of holes and electrons [198,220–224]. The existence of numerous defect states complicates the investigation of charge transport mechanisms in PZT.

$$Pb^X_{Pb} + O^X_O \rightarrow V''_{Pb} + V^{\bullet\bullet}_O + PbO(g) \qquad (5)$$

$$2[V''_{Pb}] + [e'] = 2[V^{\bullet\bullet}_O] + [h^\bullet] \qquad (6)$$

This leads to a partial Schottky equilibrium constant of

$$K_s = K_s(T) = [V''_{Pb}][V^{\bullet\bullet}_O]a_{PbO} \qquad (7)$$

where $a_{PbO}$ is the activity of PbO. Additional equilibria involve the intrinsic reaction for creation of electrons and holes represented by:

$$nil \leftrightarrow e' + h^\bullet \qquad (8)$$

$$K_i = np \qquad (9)$$

and ionization of lead vacancies given by:

$$V'_{Pb} \leftrightarrow V''_{Pb} + h^\bullet \qquad (10)$$

$$K_{ion} = \frac{[V''_{Pb}]p}{[V'_{Pb}]} \qquad (11)$$

The concentrations of defects as a function of the oxygen activity can be attained by concurrently solving the equilibrium constant equations; the result is represented via Brouwer

diagrams

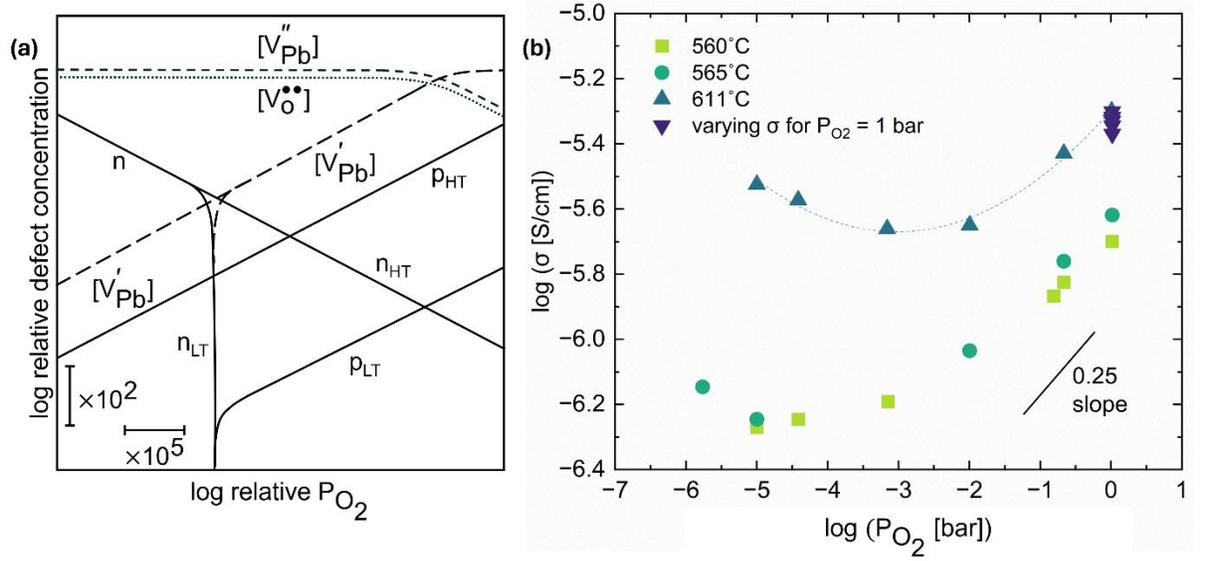

Figure 5) [207].

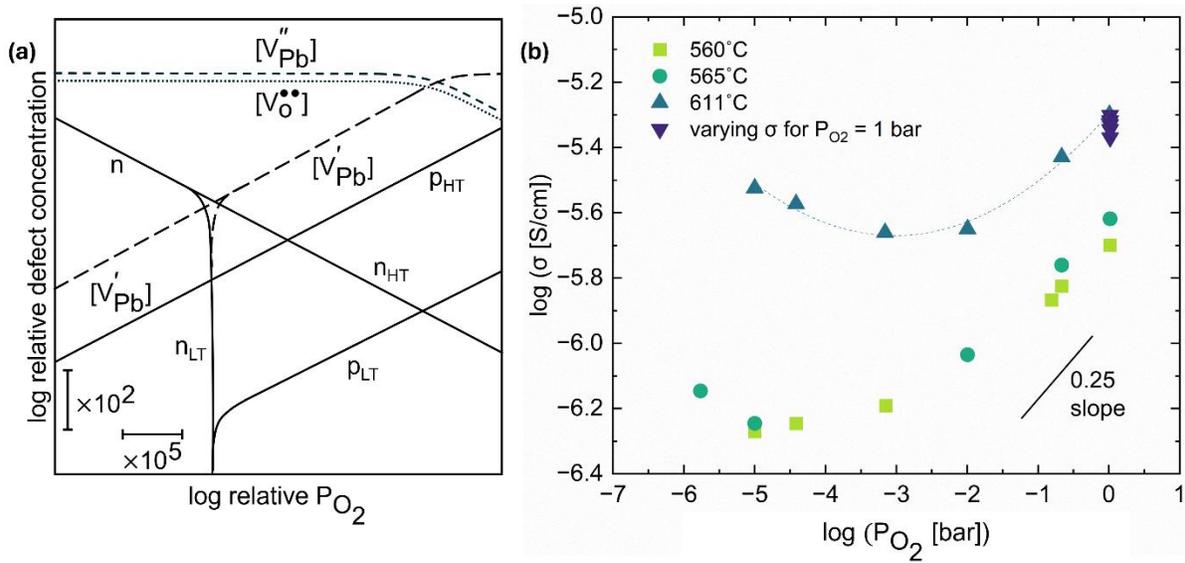

Figure 5. Schematic diagram of the defect concentrations as a function of oxygen partial pressure for PZT in the presence of deep acceptor levels, ($V_{Pb}''$, $V_{Pb}'$) [207]. (b) The dependence of conductivity on oxygen partial pressure at 560, 565, and 611°C (redrawn from data in ref. [201]).

*Reduction regime*: At oxygen partial pressures lower than those that correspond to the stoichiometric composition, the oxide will release oxygen to maintain thermodynamic equilibrium with the surrounding environment [52,102,199,201,207]. Oxygen will migrate out of the lattice as gas, each leaving behind two electrons, which can be represented as follows:

$$O_O^X \rightarrow \frac{1}{2}O_2(g) + V_O^{\bullet\bullet} + 2e' \tag{12}$$

$$K_r = P_{O_2}^{\frac{1}{2}}[V_O^{\bullet\bullet}][e']^2/[O_O^X] \tag{13}$$

As $P_{O_2}$ decreases, the electron concentration rises. These electrons can reduce $Ti^{4+}$ ions to $Ti^{3+}$, inducing polaron conductivity. The $P_{O_2}$ dependence of conductivity for acceptor doped PZT between 500 and 700°C has a slope of approximately 0.25, across a $P_{O_2}$ range from 1 to $10^{-4}$ bar; this decreases to approximately -1/6 in donor-doped PZT ceramics between 560-611°C [201]. The enhanced conductivity at lower $P_{O_2}$ is attributed to n-type electron conduction, mainly due to electron trapping by $Ti^{4+}$ [207,225].

*Oxidation regime:* As $P_{O_2}$ rises, a surface exchange reaction is initiated that consumes oxygen vacancies and generates holes.

$$\frac{1}{2}O_2(g) \to O_O^X + V_{Pb}'' + 2h^\bullet \tag{14}$$

$$K_o = [p]^2[O_O^X][V_{Pb}'']/P_{O_2}^{\frac{1}{2}} \tag{15}$$

and thus, hole conduction can dominate and the PZT films become slightly p-type even when donor doped. The conductivity of Nd doped PZT is positively correlated with oxygen partial pressures, with a slope of ¼, which can be attributed to trapping of holes via negatively charged lead vacancies. The presence of hole conduction in donor doped PZT can be accounted for by the volatilization of PbO during the fabrication process [222–224]. The consequent generation of lead vacancies alters the nominally donor-doped PZT to an oxide that, in effect, is mildly acceptor-doped. With rising temperatures, PZT exhibits increased p-type characteristics, a change that can be linked to the enhanced ionization of trapped holes [136,226,227]. These holes undergo thermally induced hopping between $Pb^{2+}$ ions [212].

Nonaka et al. investigated the influence of Pb-deficiency on the conductivity of $Pb_{1-x}Zr_{0.5}Ti_{0.5}O_{3-x}$. A significant increase in conductivity was observed in Pb deficient PZT films [228]. Furthermore, Boukamp et al. found that both ionic and electronic conductivity are strongly influenced both thermal treatment and the oxygen partial pressure; the lead vacancies are compensated by oxygen vacancies and holes trapped on the lead sites, resulting in hopping conduction for PZT samples annealed in $N_2$ [199,229]. Slouka et al. also demonstrated that PZT shows varying conductive behavior under different oxygen partial pressures. At high oxygen partial pressures, the material exhibits hole conduction, while a reducing atmosphere leads to predominant electron conduction [201]. However, the correlation between PbO stoichiometry and electrical conductivity of PZT films at different $P_{O_2}$s has not been investigated in detail in PZT films. Recent data suggests that PbO depletion from the surface of the PZT layers leads to the development of space charge regions between layers in sol-gel derived PZT films, which subsequently serve to inhibit the migration of oxygen vacancies [6]. Additionally, this loss of PbO results in the heterogeneous distribution of lead and oxygen vacancies throughout the film, which contributes to a polarity-dependent lifetime and asymmetric leakage current characteristics [230].

### 3.2.3 Donor Doping of PZT:

Undoped PZT ceramics exhibit p-type conductivity attributed to presence of unintentionally introduced acceptor ions and lead vacancies arising from PbO volatility at high temperatures, both of which necessitate either oxygen vacancies or holes to preserve charge neutrality [63,214,215,231]:

$$2[V_O^{\bullet\bullet}] + [h^\bullet] = 2[V_{Pb}''] + [A'_{Ti,Zr}] \tag{16}$$

where $V_{Pb}''$ is a lead vacancy. $h^\bullet$ denotes a hole and $A'_{Ti,Zr}$ is an acceptor impurity. The rise in the concentration of mobile $V_O^{\bullet\bullet}$ and $h^\bullet$ enhance the ionic conductivity and electrical conduction in PZT, respectively [42,200,216].

$$2[V_O^{\bullet\bullet}] + [Nb_{Ti}^\bullet] + [h^\bullet] = 2[V_{Pb}''] + [e'] \tag{17}$$

When donor ions like La and Nd are added to the A-site, or Nb, Sb, and Ta to the B-site, charge compensation can occur either ionically through the formation of cation vacancies or electronically through the creation of holes. Thus, Nb-doped PZT reduces the concentrations of oxygen vacancies and holes [46,62,63,136,200,212,230–245]. However, completely eradicating oxygen vacancies in donor doped PZT ceramics can be challenging, as it requires first compensating for any PbO deficit. An approximately 2 mol% PbO loss upon annealing PZT ceramic at 650°C was reported [245]. Non-stoichiometry in PZT due to PbO loss can be prevented by (1) firing the ceramic with a lead source (2) addition of extra PbO to the film solution or/and utilization of PbO excess target during sputtering, and (3) post-annealing under PbO rich atmosphere [246]. In cases, where no excess PbO is added, it is reasonable to anticipate the occurrence of p-type conductivity in PZT ceramics, including those moderately doped with donor ions, nor does donor doping completely inhibit the generation of oxygen vacancies [200,233,236]. Enhancements in the resistance to leakage currents have been achieved through the incorporation of donor dopants at both the A-site and B-site in the PZT. The improvement in leakage resistance was attributed to decreases in the hole and oxygen vacancy concentrations, with this improvement tapering off at lower temperatures. This trend is likely linked to low equilibrium concentration of free holes. As the temperature is lowered, the insulation resistance appears to plateau, potentially influenced by charge injection phenomena [240].

Oxygen vacancy migration in doped PZT was examined using combined impedance spectroscopy and oxygen tracer experiments. Nd-doped PZT exhibits significant ionic conductivity above 600°C, mainly due to enhanced oxygen vacancy migration at grain boundaries, complemented by measurable but lower oxygen vacancy mobility in the bulk material (Figure 6) [201]. $^{18}$O tracer studies revealed increased concentrations at grain boundaries (2 ppm), indicating higher vacancy concentrations there compared to the bulk (0.01 ppm) [201,247].

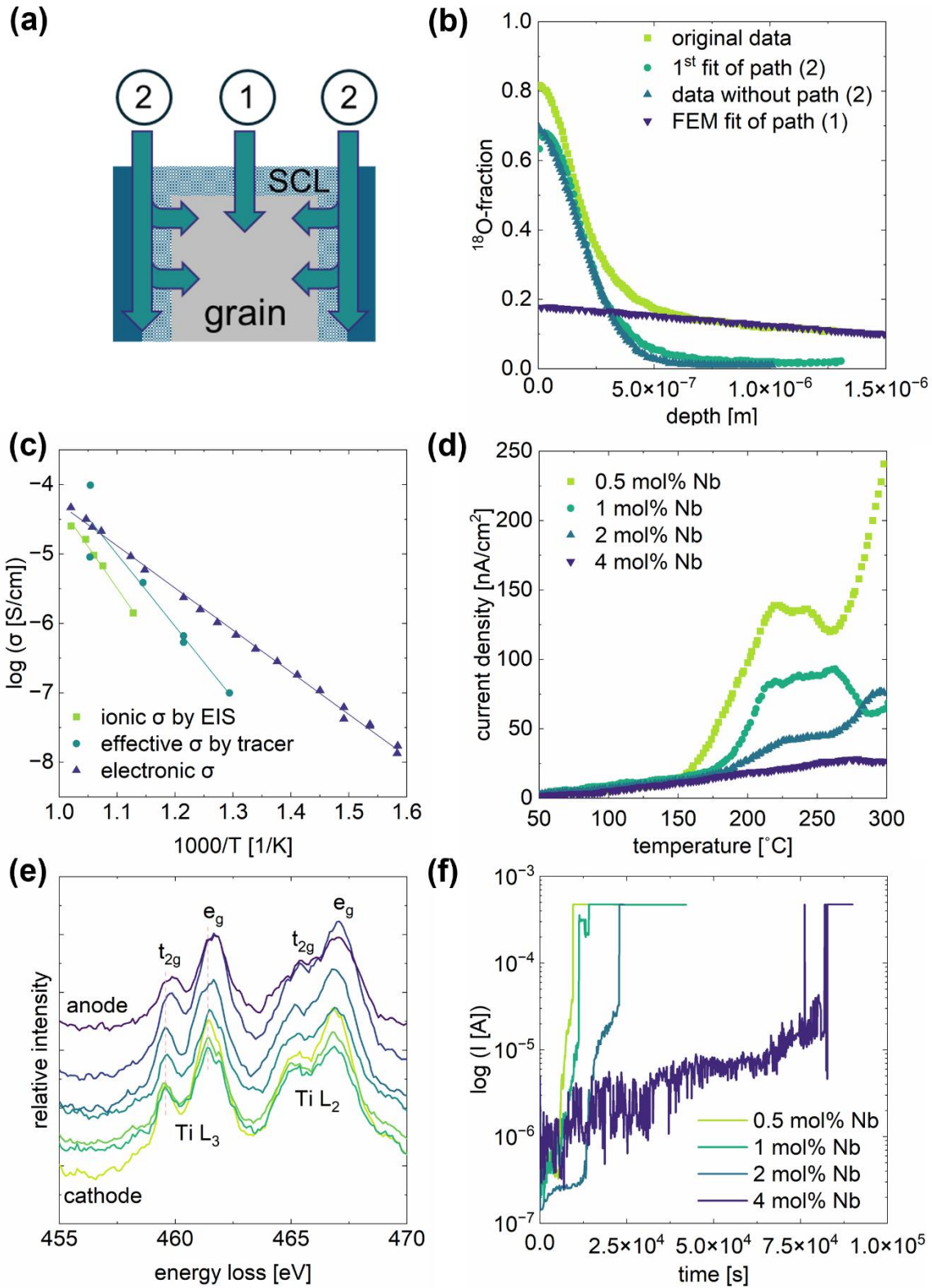

Figure 6. (a, b) The model for analyzing tracer diffusion profiles incorporates two distinct diffusion pathways. Path (1) encompasses a space charge layer (SCL) near the surface. Path (2) represents grain-boundary transport, which may occur either in the core of the grain boundary or along a space charge layer where oxygen vacancies accumulate

[248]. (c) Arrhenius plot of conductivities estimated from impedance spectroscopy together with the effective ionic conductivity calculated from $D_{gb}$, (d) Evolution of TSDC peaks in degraded PZT films with increasing Nb concentration (redrawn from data in ref. [249]), (e) Electron energy loss spectroscopy profiles at varying positions in the 2 mol% Nb doped PZT film, mapped relative to the distance from the cathode interface, where oxygen vacancies accumulate upon electrical degradation. Ti $L_{2,3}$ of 2 mol% Nb doped PZT films after degrading at 180°C under electric field of 350 kV/cm for 12h, (f) Change in leakage current as a function of time for a film exposed to 350 kV/cm at 180°C (redrawn from data in ref.[250]).

The conductivity changes induced by donor ions in PZT thin films have been a subject of extensive research [46,175,230,237,240,250,251]. Notably, the defect chemistry in thin films diverges from that in bulk ceramics, which in turn influences the charge transport mechanisms in thin films. Three of the critical disparities are: (1) Thin film processing often occurs away from thermodynamic equilibrium, potentially leading to elevated point defect concentrations and the formation of metastable structures; (2) Interfaces exert a pronounced effect on the point defect chemistry in films compared to bulk materials, with the equilibrium concentration of defects being heavily reliant on their formation energy at the interface; and (3) As the feature size decreases, the required diffusion distance for defects to contribute to degradation also diminishes. Mihara et al. revealed that the oxygen vacancy concentration near the surface ($10^{20}$ cm$^{-3}$) is higher than that in the bulk of PZT films ($10^{18}$ cm$^{-3}$) [252]. The elevated oxygen vacancy concentration near the interface decreases the potential barrier for electron injection, owing to the Fermi level pinning nearer to the conduction band [253,254]. This results in a n-type conductivity at the PZT interface while maintaining p-type characteristics in the bulk of the PZT film [250].

The degradation of DC resistance in PZT films is markedly affected by the specific defect chemistry at the interfaces near the electrodes, which can vary between the top and bottom interfaces [230]. This variance is due to several factors: firstly, a non-uniform distribution of defects throughout the PZT films, which might include an oxygen concentration gradient caused by variations in the PbO stoichiometry; secondly, the potential Ti/Zr segregation during the film fabrication process; and thirdly, fluctuations in the oxidation states of ions within the film, which may serve as charge trap sites. Al Shareef et al. demonstrated significant asymmetry in leakage currents and breakdown voltages in Nb-doped PZT films, even with platinum electrodes on either side [238]. Several investigations have linked the electrical breakdown behavior in PZT films to the polarization-dependent characteristics of the Schottky barrier [251,255].

The interdependencies among defect chemistry, leakage currents, and time dependent electrical degradation was explored for 0.5-4 mol% Nb doped PbZr$_{0.52}$Ti$_{0.48}$O$_3$ (PZT) films. While the films are nearly intrinsic, experimental evidence indicated hole hopping between Pb$^{2+}$ and Pb$^{3+}$ states, alongside electron capture by $Ti^{4+}$ sites.[69] For all Nb levels, the dominant electrical conduction rate-controlling mechanism shifted from Poole-Frenkel emission at lower electric fields to Schottky emission as the electric field increased. This transition field was observed to reduce with a decrease in Nb content. The accompanying decrease in Schottky barrier height, associated with lower niobium levels, is explained by the pinning of the Fermi level due to oxygen vacancy formation [230,250]. It is important to note, however, that the details of the conduction mechanism will depend explicitly on the processing-dependent defect chemistry.

The drift of oxygen vacancies and subsequent accumulation near the cathode was linked to the DC resistance degradation [45,61–63,256–259]. Notably, a rise in oxygen vacancy concentration near the cathode lowered the electron injection barrier from the cathode to the film's interior [250]. The injected electrodes then trapped via $Ti^{4+}$ to reduce $Ti^{3+}$. The existence of $Ti^{3+}$ near the cathode was verified via EELS (Figure 6.e). The decrease in oxygen vacancy concentration with Nb significantly improved time dependent leakage current characteristics (Figure 6.f) [250].

When the Nb concentrations in the PZT films exceed 6 mol%, a profound transformation in defect chemistry and the associated charge transport dynamics was observed. At these higher doping levels, several phenomena were identified: firstly, the superoxidation at the surface of the PZT films resulted in a more p-type character at the PZT/Pt interface, which in turn diminished electron injection across the Schottky barrier. Secondly, within the bulk of the film, the dominant charge transport process transitioned from electron trapping at $Ti^{4+}$ sites to hole conduction between lead vacancies. Lastly, there was a notable reduction in ionic conductivity attributed to the diminished migration of oxygen vacancies. Despite a profound change in defect chemistry contributed to electronic and ionic conduction in heavily Nb doped PZT films, a small variation in electrical lifetime was observed [260,261].

### 3.2.4 Acceptor Doping of PZT:

Acceptor type impurities like $Fe^{2+,3+}$, $Mn^{2+,3+}$, $Mg^{2+}$ have lower valence than the B-site cations ($Ti^{4+}$, $Zr^{4+}$) they replace. Charge compensation is achieved by generating oxygen vacancies and/or holes; in some cases, these point defects associate to create defect dipoles such as $(A_{Ti}'' - V_O^{\bullet\bullet})^x$ or $(A_{Ti}' - V_O^{\bullet\bullet})'$ [262]. Electroneutrality in acceptor doped PZT can be described by:

$$2[A_{Ti}''] + [A_{Ti}'] + 2[V_{Pb}''] + [e'] = 2[V_O^{\bullet\bullet}] + [h^\bullet] \qquad (18)$$

Multivalent ions such as $Mn^{2+/3+/4+}$ or $Fe^{2+/3+/4+}$ enrich the charge transport properties of PZT by acting as electron and hole traps. As these ions switch between their oxidation states, they capture or release charge carriers, affecting the overall electrical conductivity [263,264].

$$A'_{Ti} + e' \rightleftarrows A''_{Ti} \qquad A^x_{Ti} + e' \rightleftarrows A'_{Ti} \qquad (19)$$

$$A''_{Ti} + h^\bullet \rightleftarrows A'_{Ti} \qquad A'_{Ti} + h^\bullet \rightleftarrows A^x_{Ti} \qquad (20)$$

Oxygen tracer diffusion in acceptor ($Fe^{3+}$)-doped PZT revealed that acceptor doping increases the amount of oxygen vacancies by a factor of 400 relative to a 1 mol% La doped PZT ceramic. This leads to a higher bulk vacancy diffusion coefficient and ionic conductivity. Additionally, it has been suggested that the electrical conductivity is influenced by the trapping of holes by $Pb^{2+,3+,4+}$ and $Fe^{2+,3+,4+}$ ions. It appears that trapped holes play a significant role in counterbalancing the charges introduced by the acceptor dopants [200]. The study found that $Fe^{3+}$ doped PZT has higher diffusion rates for oxygen vacancy migration within the bulk material compared to $La^{3+}$ doped PZT, without any signs of accelerated diffusion along the grain boundaries [200].

Likewise, in acceptor doped PZT films, there are indications of electronic conductivity resulting from charge trapping by acceptor ions and ionic conductivity due to the movement of oxygen vacancies [164,265–270]. While charge injection from a Pt electrode into Mn doped PZT films is controlled by Schottky emission, the drift of injected charges is dominated by Poole-Frenkel conduction via carrier hopping between redox active Mn trap sites ($Mn^{2+}$, $Mn^{3+}$, $Mn^{4+}$) [268–270]. In Mn doped PZT films, electromigration and subsequent accumulation of oxygen vacancy migration was monitored by TSDC (Figure 7.a). In principle, each oxygen vacancy is compensated by two electrons ($n \sim 2[V^{\bullet\bullet}_O]$), which can be trapped by the $Ti^{4+}$ to create $Ti^{3+}$. However, no apparent chemical shift is observed from a series of EELS spectra taken across the thickness of degraded Mn-doped PZT, implying that there is little modification to the Ti valence (Figure 7.b). This suggests that despite the higher oxygen vacancy concentration induced in PZT films by acceptor ions, the generation of free electrons through oxygen vacancy compensation at the cathode and the formation of free holes within the anode region may be restrained by the valence transitions from $Mn^{3+/4+}$ to $Mn^{2+/3+}$ and $Mn^{2+/3+}$ to $Mn^{3+/4+}$, respectively (Figure 7.c). Due to the lower concentration of Mn with respect to Ti in PZT films, carriers trapped on Mn sites are distributed over a longer length scale. This diminishes the maximum electric field at the Schottky interface, which suppresses the potential barrier height lowering and subsequent leakage current rise upon degradation. This leads to longer lifetimes and lower electrical degradation rates in Mn doped PZT films [269]. Furthermore, when a layer of acceptor-doped PZT is introduced at the interface between the cathode and a donor-doped PZT stack, an improvement in the film's resistance to electrical degradation is noted [270].

The contribution of electron/hole trapping to electrical degradation is strongly dependent on the relative concentrations of $A_{Ti}^{X}, A_{Ti}'$, and $A_{Ti}''$, which are influenced by the annealing atmosphere. Hayashi et al. reported that the concentration of $Mn^{4+}$ ions significantly decreased after annealing $PbTiO_3$ in a reducing atmosphere [41,271,272]. Similarly, it was found that the $Mn^{2+/3+}$ ions in PZT films are ionized to $Mn^{3+/4+}$ ions by absorbing oxygen from the atmosphere upon heating at 320°C. The decrease in $Mn^{2+/3+}$ with heating was demonstrated using Q-DLTS (Figure 7.f)) [41]. The electrical conductivity of 2% Mn doped PZT film increased approximately two orders of magnitude after holding at 320°C for 366 h (Figure 7.d). The variation in the valence state of Mn ions through the thickness of PZT films contributes to the polarity dependent electrical degradation and lifetimes [41].

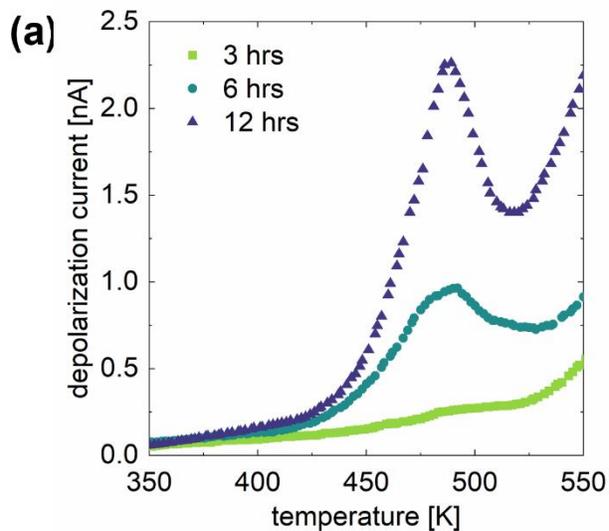
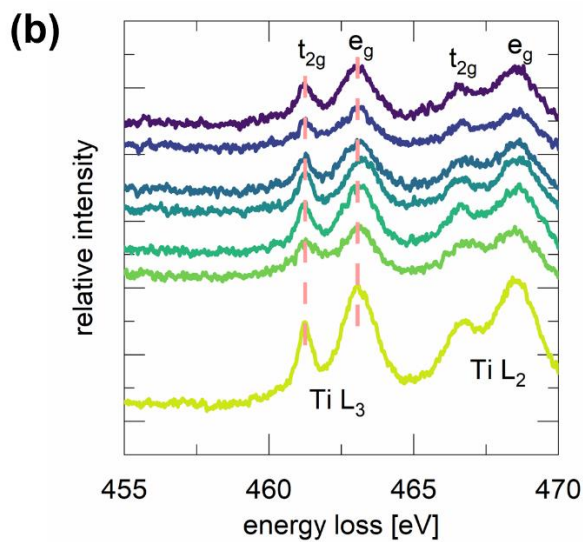
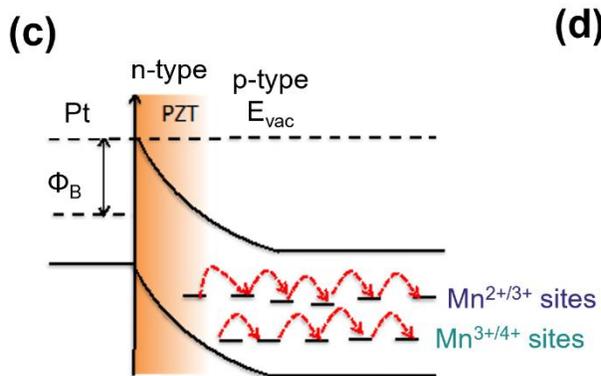
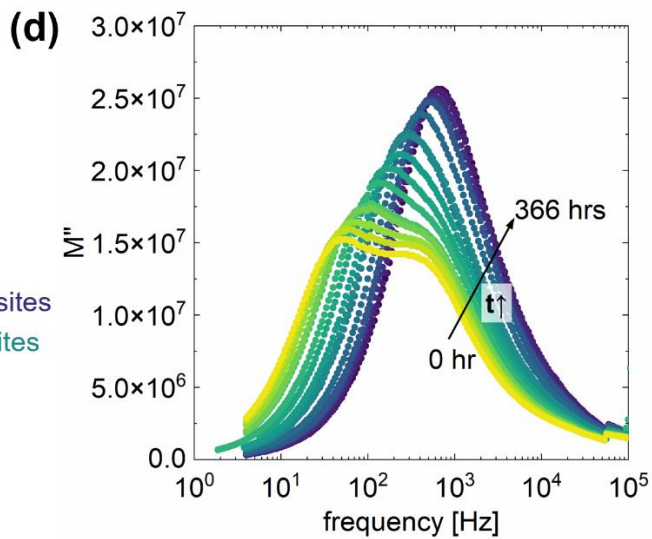
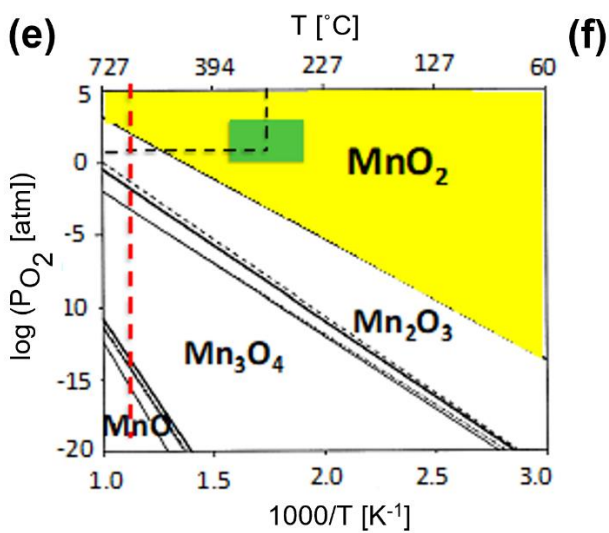
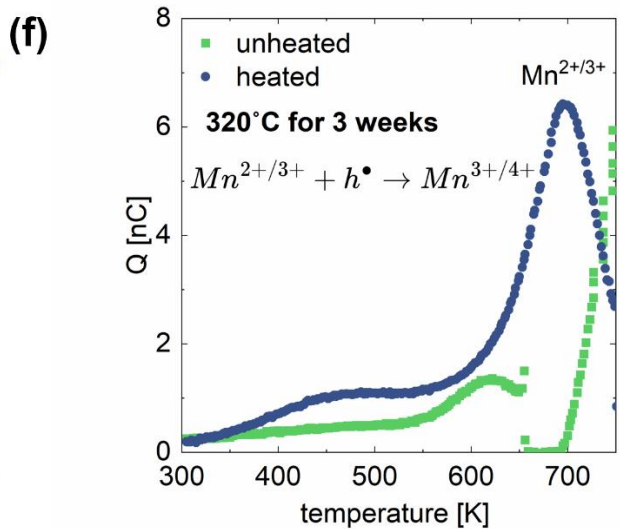

Figure 7. (a) TSDC spectra of 2% Mn doped PZT films upon degradation at 180°C under an electric field of 350 kV/ cm for 3, 6, or 12 hrs, (b) Electron energy loss spectra of electrically degraded 2 mol% Mn doped PZT film as a function of distance from the cathode (redrawn from data in ref. [269]), (c) Schematic representation of Mn impurity sites in PZT energy band diagram (reproduced with permission from ref. [269]). (d) Imaginary modulus of 2% Mn doped PZT plotted as a function of frequency after heating at 320°C for different times, (e) Mn-O phase stability diagram (reproduced with permission from ref. [273]), and (f) Q-DLTS of 2% Mn doped PZT films before and after heating at 320°C for 3 weeks (redrawn from data in ref. [41]).

## 4 Probing Defect Chemistry: Techniques and Insights into Degradation Assessment

A variety of characterization techniques are employed to gain insights into defect chemistry, with each offering differing perspectives, as shown in Figure 8. However, no single method can entirely encapsulate the wide spectrum of defects, necessitating a comprehensive approach that combines spectroscopic, microscopic, and electronic techniques.

In this section, characterization techniques are categorized in three sections. The first group focuses on local analysis of composition, microstructure, and bonding, crucial for identifying the defects responsible for electrical degradation. A second set of methods centers on the detection of trap levels, which are potential sources for releasing or trapping charges; excessive released charge can potentially initiate electrical breakdown via thermal breakdown processes. Thirdly, there are techniques dedicated to characterizing degradation. These methods investigate and analyze how defects impact the overall stability and performance of the materials. Together, these approaches, with their distinct temporal and spatial resolutions, offer a comprehensive view of defect concentrations, structures, and mobility.

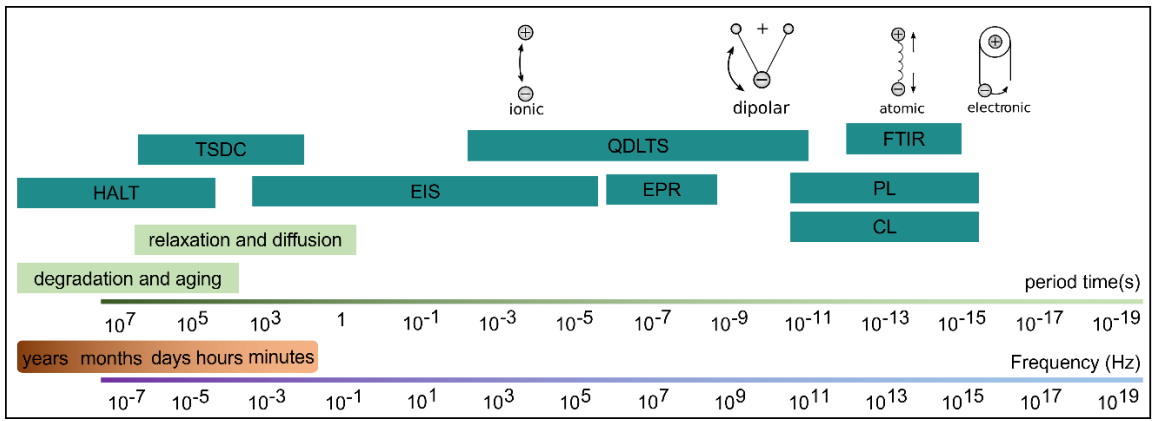

Figure 8. Classifying defect characterization techniques according to relaxation time scales.

### 4.1 Local Analysis of Composition and Bonding: Connecting Localized Probes to Defect Chemistry

#### 4.1.1 Time-of-flight Secondary Ion Mass spectrometry (TOF-SIMS)

Time-of-Flight Secondary Ion Mass Spectrometry (TOF-SIMS) provides critical insights

into the electrical degradation of dielectric and piezoelectric materials. Traditional SIMS approaches, which rely on electrostatic or magnetic spectrometers, suffer from low transmittance due to the necessity of narrow slits that only permit ions with the specific mass/charge ratio to pass through. TOF-SIMS overcomes these limitations by employing a pulsed ion beam with beam diameters as small as 0.3 µm, commonly generated by a liquid $Ga^+$ gun, which allows for a more efficient collection of ions and hence, an increase in ion transmittance by 10–50% [274]. This method also significantly reduces the sputtering rate, making it ideal for characterizing organic surface layers and detecting a wide range of ion fragments. The sensitivity of TOF-SIMS to surface metals is particularly noteworthy, with the ability to detect surface densities as low as $10^8$ atoms/$cm^2$ for elements like Fe, Cr, and Ni on Si surfaces [275]. TOF-SIMS enables dopant profiling in semiconductors, where its utility exceeds that of traditional SIMS due to minimal matrix effects and linear ion yield proportionality up to 1%.

ToF-SIMS is a critical instrument in exploring electrical degradation mechanisms associated with oxygen profiles in oxides [88,276–282]. For example, Rodewald et al. [142] utilized SIMS to examine oxygen isotope exchange in Fe-doped STO single crystals during electrocoloration. Their study found that Au/Cr electrodes almost entirely blocked ion movement, impeding oxygen exchange, whereas Ag/Cr electrodes demonstrated some degree of oxygen incorporation, suggesting partial ion permeability. De Souza et al. [279] utilized ToF-SIMS to simultaneously detect multiple secondary ions for comprehensive analysis within a single SIMS depth profile. As a result, ToF-SIMS provides details about oxygen isotope profiles, dopant accumulation, and ion-beam mixing, even within 5 nm of the surface of Fe-doped STO samples. Using ToF-SIMS, De Souza et al. [280] established that oxygen vacancy concentrations in undoped STO crystals largely remain unchanged after post-oxygen annealing. The isotope diffusion profiles revealed that the depletion space-charge layers on STO surfaces are predominantly driven by surface-level oxygen vacancy creation.

In a complementary study, Frömling et al. [276] used $^{18}O$ tracer ToF-SIMS depth profiles to explore oxygen diffusion, particularly along grain boundaries in donor doped BTO. Rapid grain boundary diffusion was shown to occur between 750°C and 900°C. They also identified oxygen vacancy-enriched space-charge regions which influence oxygen grain boundary diffusion, in contrast to Ti-rich secondary phases which had a minimal impact on oxygen transport in BTO. In addition, Opitz et al. [184] used $^{18}O$ to study oxygen diffusion in BTO-based MLCCs, revealing fast diffusion along the Ni-BTO interface. These regions often have a higher oxygen vacancy concentration during sintering, making them the natural pathways for oxygen during the reoxidation stage. Following this pathway, the oxygen then progresses along grain boundaries and subsequently permeates into the grains of the BTO. This observation highlighted the critical role of interfaces and grain boundaries in the diffusion process of composite materials.

### 4.1.2 Electron Microscopy Analysis: Electron Energy-Loss Spectroscopy and EBIC Imaging

High Resolution Transmission Electron Microscopy (HRTEM) and Electron Energy-Loss Spectroscopy (EELS) are essential techniques for probing defect chemistry and degradation mechanisms. HRTEM provides high-resolution imaging of atomic structures and defects, while EELS analyzes electron energy distributions after transmission through a material. EELS is particularly sensitive to low atomic number elements ($Z \leq 10$), complementing Energy Dispersive X-ray Spectroscopy (EDS) which detects elements with $Z > 10$ [274]. Integrating advanced techniques such as atomic-resolution scanning transmission electron microscopy (STEM), HRTEM, and EELS facilitates study of structure-property relationships at the atomic level. These methods combine in-depth analysis of chemical composition and bonding, with the ability to probe crystals at sub-angstrom resolution. They are particularly effective for identifying microscopic defects and disorder in thin films and interfaces. EELS, integrated with STEM, offers elemental and bonding information, particularly through the analysis of O-K edge and transition metal $L_{2,3}$ edges. These techniques are invaluable for examining electrically active impurities and defects, offering insights into structural and electrical properties at the micro and nano scales, and revealing changes in material composition during degradation processes.

The microstructure of perovskites is often studied using HRTEM [283–287]. HRTEM allows cation columns to be imaged along the viewing direction, leveraging the high nuclear charge of cations for strong phase contrast due to their substantial scattering power. Conversely, imaging the oxygen sublattice is more challenging due to its relatively low scattering power. Jia et al. [285] developed a novel imaging mode by modifying the spherical-aberration coefficient of the objective lens in a TEM. This adjustment enabled comprehensive imaging of *all* atomic columns in STO, including the oxygen. Visualization of the oxygen facilitates both detection of local nonstoichiometry and the extent of oxygen-vacancy ordering. For example, Muller et al. [284] used HRTEM and EELS to study the distribution of oxygen vacancies in $SrTiO_{3-\delta}$ films, focusing on changes after degradation. EELS served as an indirect method to detect oxygen vacancies. They observed that the shape of the EEL spectra exhibited features consistent with the presence of oxygen vacancies and achieved a detection limit for $V_O^{\bullet\bullet}$ of ~0.05 (~1 at% $V_O^{\bullet\bullet}$), determined by the ratio of $Ti^{3+}/Ti^{4+}$ calculated from the EEL spectra in STO crystals.

The electronic properties of dielectric and piezoelectric are highly sensitive to the composition and atomic structure of the grain boundaries [153,277,288–292]. The macroscopic electrical properties of STO have often been explained through the concept of double Schottky barriers, which are thought to arise from charged grain boundaries with compensating space charge in depletion layers. This results in an electrostatic potential that hinders the movement of free carriers across the grain boundary. Kim et al. [153] utilized a combination of Z-contrast imaging, EELS, and first-principles theory to study grain boundary dislocation cores in STO, revealing that these cores are intrinsically nonstoichiometric, with a higher Ti:O ratio than the bulk material. Z-contrast imaging provided the geometric structure of the boundaries, while EELS confirmed the nonstoichiometry. Theoretical analysis identified the most energetically favorable atomic arrangements and corroborated the findings from imaging and EELS. This comprehensive approach concluded that Ti-core boundaries in STO have excess titanium, whereas Sr-core boundaries are oxygen deficient, both contributing to excess electrons at the boundary plane. The calculations also revealed that these nonstoichiometric grain boundaries contribute to what is conventionally described as a double Schottky barrier, but which might more accurately be characterized as a p-n-p double junction at the grain boundary.

HRTEM and EELS also facilitate the investigation of oxygen vacancy diffusion and ordering to create clusters in BTO, which play a significant role in the degradation of insulation resistance. Yang et al. [76] examined planar defects in BTO grains in degraded BME MLCCs, as shown in Figure 9. These defects result from clustering of oxygen vacancies in the perovskite lattice, altering the standard lattice composition to $Ba - O_{3-x}[V_O^{\bullet\bullet}]_x$ (0 < x < 3) instead of Ba–$O_3$ on the [111] planes, which causes local lattice distortions. The high concentration of oxygen vacancies in the degraded dielectrics leads to a reduction in the valence of Ti, as confirmed by EEL spectra.

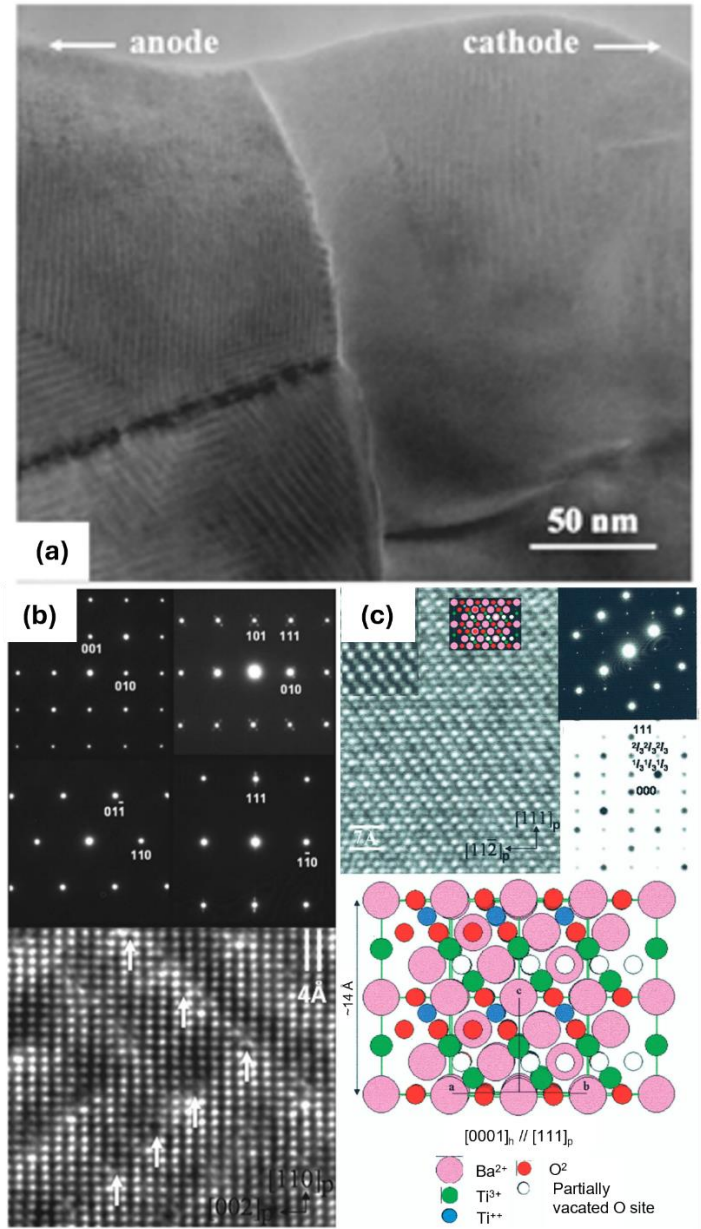

Figure 9. (a) Bright-field image displaying typical distributions of modulated defect within individual BTO grains in the degraded capacitors (reproduced with permission from ref. [82]), (b) electron microscopy data from degraded $BaTiO_{3-\delta}$ with incommensurate modulations, (c) degraded $BaTiO_{3-\delta}$ with an ordered structure of oxygen vacancies (reproduced with permission from ref. [80]).

The voltage contrast technique, frequently used in Focused Ion Beams (FIB) and Scanning Electron Microscopy (SEM), is another method for gaining insights into defects in semiconductors. In MLCCs, this method can identify locally failed or highly degraded regions. It operates on the principle that secondary electron emission from a surface, when excited by a primary electron beam, differs based on the surface's electrical charge. Failures are typically visible as variations in brightness [293–295]. Yang et al. [296] applied this technique in a dual beam FIB/SEM to investigate heterogeneous defects in Ni–BTO MLCC at a submicron level. They observed that the Ni layers, depending on their electrical polarity, appeared either darker or brighter. Notably, the brighter areas extended across the dielectric layers near the internal electrode ends, as shown in Figure 10. This was linked to a high concentrations of oxygen vacancies in the BTO, as confirmed by EELS analysis. EELS data showed significant changes in the fine structure of Ti $L_{2,3}$ and O K edges between semiconducting areas (spectrum 1) and pristine BTO regions (spectrum 2). The Ti $L_{2,3}$ edges, which were split in spectrum 2, showed no such splitting in the conductive regions. This transition is triggered by the reduction from $Ti^{4+}$ to $Ti^{3+}$, suggesting a high oxygen deficiency in these areas. High-resolution TEM/EELS analysis indicated that the high oxygen vacancy concentration created conduction paths through electron hopping from $Ti^{4+}$ to $Ti^{3+}$, significantly reducing insulation resistance in these areas.

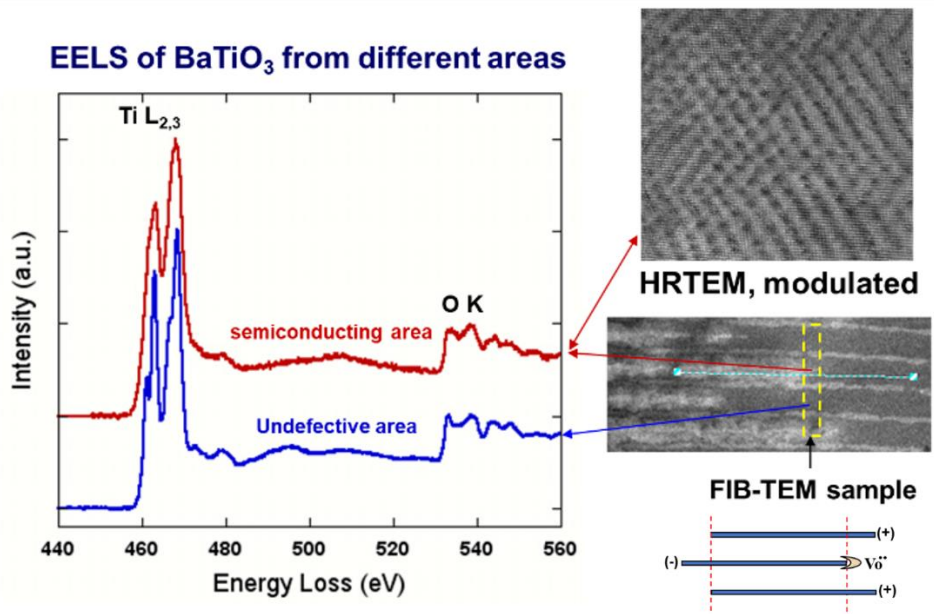

Figure 10. Voltage contrast SEM image obtained from a degraded Ni–BTO MLCC with corresponding EELS spectrum obtained from the semiconducting area and a pristine BTO region, and HRTEM image of semiconducting area demonstrating defect-induced modulation within BTO grains in the degraded MLCC (reproduced with permission from ref. [296]).

Electron Beam Induced Current (EBIC) is another technique for studying defect chemistry and material degradation. It detects electrically active impurities by harnessing minority carriers generated in a junction device with an electron beam. Its most typical form involves generating current through the field-separation of electron-hole pairs created by the beam. EBIC maps local electric fields and carrier dynamics in electronic devices, identifying defects like dislocations and grain boundaries for semiconductor dopant density analysis [196]. The non-contact nature of EBIC allows for detailed examination of small areas. Additionally, EBIC can profile the minority carrier diffusion length and lifetime [274]. Another variant, Electron Beam Absorbed Current (EBAC), measures the absorbed electron beam current to quantify electrical connectivity in circuits. Both these modes are predominantly used in SEM, with occasional applications in STEM for enhanced resolution [194,197–199]. Secondary electron (SE) emission EBIC (SEEBIC), a more recent development in EBIC modes, detects the current generated in a sample following the emission of Ses. SEEBIC is employed to quantitatively map resistance, potential, electric field, and work function in devices, and offers notably higher resolution than other EBIC methods. The choice of EBIC mode depends on various factors, including the circuit, device morphology, and sensitivity of current measurement. In standard EBIC, effective measurement needs one carrier of the separated electron-hole pairs to reach ground; this is often achieved by grounding one side of the circuit and connecting the other to a transimpedance amplifier. Both EBAC and SEEBIC can operate with a single sample connection and can generate resistance contrast images when an additional ground path is introduced. By analyzing the current distribution between the transimpedance amplifier and ground, these methods can map resistance at each pixel [297,298].

Hubbard et al. [299] employed EBIC in STEM to map electronic transport in a single dielectric layer of a polycrystalline BTO MLCC. SEEBIC resistance contrast imaging highlighted grain boundary resistance steps, while standard EBIC simultaneously revealed Schottky barriers in the same sample. SEEBIC also enabled the calculation of the device's potential and electric field under bias. Their analysis revealed a combination of high and low resistance, Ohmic, and Schottky barriers at grain boundaries.

### 4.1.3 X-ray Photoelectron Spectroscopy (XPS)

X-ray Photoelectron Spectroscopy (XPS) is an essential analytical tool for surface chemical analysis. It employs the photoelectric effect, where X-rays eject electrons from a material's surface, enabling the identification of chemical species and their electronic states up to a depth of about 10 nm. XPS determines the surface elements' binding energy, which is, in turn, a function of the chemical environment. Its ability to detect almost all elements except hydrogen and helium makes it critical in understanding material chemistry in the presence of defects or impurities [274]. Furthermore, XPS is instrumental in determining the Fermi level in materials. This includes analyzing how the Fermi level changes in doped materials or in response to surface chemistry and defect states. Additionally, XPS can measure Schottky barrier heights at metal-semiconductor interfaces, as shown in Figure 11.

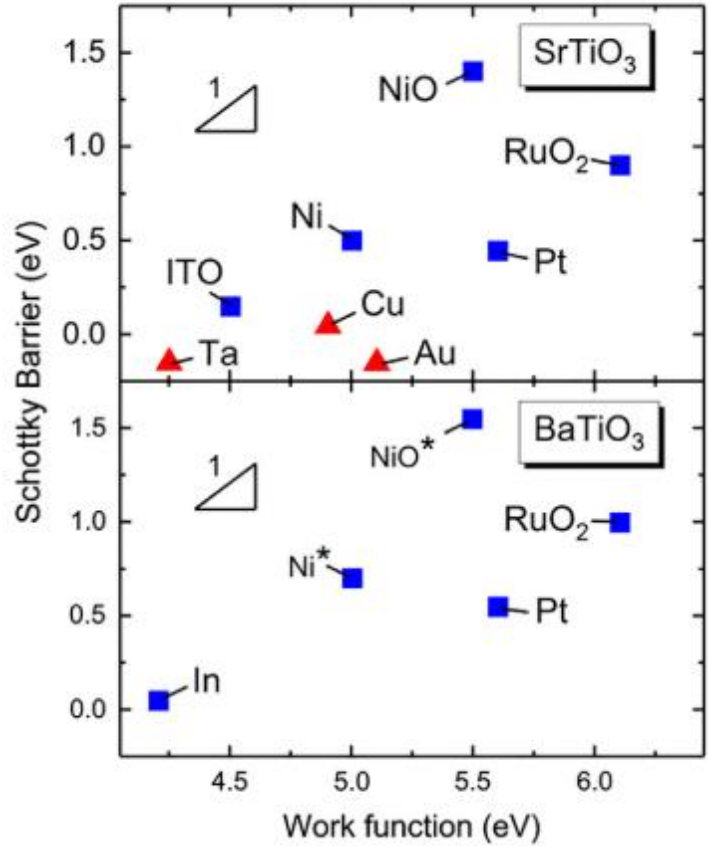

Figure 11. Schottky barrier heights for holes for different electrode materials on STO (top) and BTO (bottom). Blue squares are for nonreactive interfaces and red triangles for reactive interfaces (reproduced with permission from ref. [300]).

XPS of perovskite oxides like BTO and STO [300–309] measures core electron binding energies to reveal oxidation states, chemical environments, and electronic structures, providing critical insights into defects such as oxygen vacancies or dopants in these materials. In a recent application, Pan et al. [308] used XPS to study the reliability of MLCCs manufactured with BTO powders synthesized via solid-state (S-MLCC) and hydrothermal methods (H-MLCC). Their study revealed that S-MLCCs outperformed H-MLCCs in DC dielectric breakdown and accelerated lifetime tests. XPS analysis showed that S-MLCC had a lower $Ti^{3+}$ ion concentration and reduced Ni diffusion compared to H-MLCC, resulting in higher insulation resistance and a lower concentration of oxygen vacancies. Consequently, S-MLCC displayed a higher activation energy for conduction at both the grain and grain boundary, indicating superior performance and reliability. Moreover, Sheng et al. [303,310] observed an anomalous increase in current as a function of time when iron-doped titania ceramics were subjected to a DC electrical field. TSDC analysis revealed a peak at approximately 170°–180°C, attributed to the depolarization of oxygen vacancies. XPS further indicated that electrical stress leads to reduction within the ceramic, especially on the cathode side. These findings support the idea that the anomalous current arises from a blockage of the $O_2(g)/O^{2-}(s)$ ion transfer at the cathode, leading to the formation of a space charge in that region.

Klein et al. [254,300,311–316] extensively researched the chemical and electronic properties of dielectric oxide interfaces, focusing on the effects of various fabrication and treatment techniques. For instance, Hubmann et al. [317] utilized *in-situ* XPS to study the relationship between ferroelectric polarization and Schottky barrier height at $BTO/RuO_2$ interfaces. Using BTO single crystals with different orientations, they showed that Schottky barrier heights in unpoled samples were consistent across different crystal orientations, aligning with previous research on polycrystalline PZT [313].

### 4.1.4 Kelvin Probe Force Microscopy (KPFM)

Kelvin Probe Force Microscopy (KPFM) analyzes the local surface potential in materials, such that the electric field distributions under DC bias conditions can be quantified [318–321]. The precision of KPFM in mapping electric fields greatly enhancing fundamental understanding of degradation mechanisms [322–328].

In a groundbreaking study by Okamoto et al. [329–332], KPFM was utilized to investigate insulation degradation mechanisms in BME MLCCs with Ni electrodes. These authors noted that in the early stages of degradation, the local electric field intensified near the cathode. At the intermediate stage, a homogeneous field was observed, and in the final stage, the field strength shifted towards the anode side in the MLCCs undergoing degradation, as shown in Figure 12. This shift suggests that degradation initiates from an increase in hole density near the anode due to the electromigration of oxygen vacancies, followed by a continuous rise in hole density across the component. This effect is particularly significant in MLCCs with Ni-Sn internal electrodes on the anode side [333,334]. Leakage current analyses in degraded BME MLCCs revealed distinct current conduction mechanisms: tunnelling for Ni internal electrodes and Schottky or Poole–Frenkel for Ni–Sn internal electrodes. Additionally, HRTEM identified a significant Sn-rich interface layer at the junction between BTO and Ni-Sn internal electrodes. This layer is vital in mitigating leakage current degradation in MLCCs, highlighting the importance of electrode composition and interface structure in influencing the performance and longevity of these components [333].

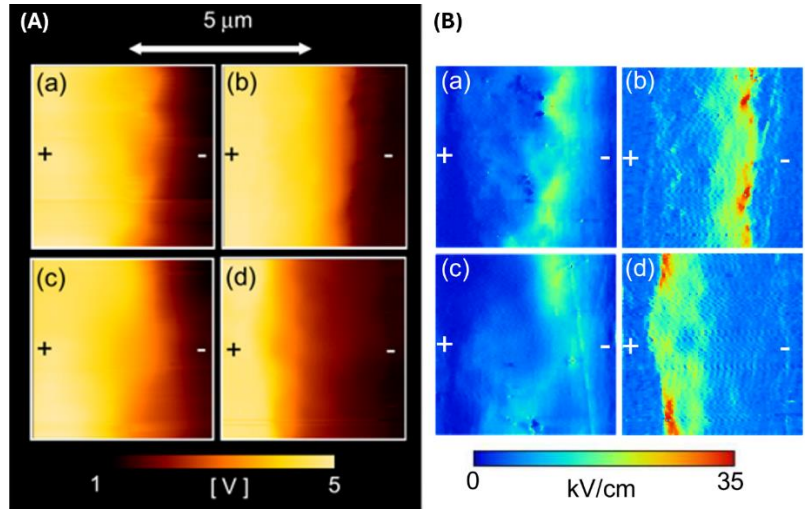

Figure 12. Electric (A) potential and (B) field distribution maps of degraded BTO-based X5R MLCC under 5 V bias voltage for (a) 1 hr, (b) 10 hrs, (c) 30 hrs, and (d) 100 hrs (reproduced with permission from ref. [329]).

Zhang et al. [335] employed the Finite Element Method (FEM) and KPFM to assess the impact of electrode defects on the electrical reliability of MLCCs. Their research focused on visualizing electric field concentration around electrode defects. It was found that MLCCs with superior electrode quality exhibited higher reliability, a conclusion that aligns with previous findings by Samantaray [128,336,337]. Moreover, Du et al. [338] utilized the same techniques to analyze local electric field concentration around pores in MLCCs. It was found that electric fields concentrate near dielectric, interface, and electrode pores, with the pore's geometric orientation playing a crucial role in local field strength. Pores contribute to insulation degradation, serving as centers for local electric and thermal stress and oxygen vacancy accumulation.

### 4.1.5 Electron Paramagnetic Resonance (EPR)

Electron Paramagnetic Resonance (EPR) spectroscopy is a highly sensitive and high-resolution technique, primarily employed for identifying and characterizing point defects in paramagnetic charge states. EPR spectroscopy utilizes microwave photons to induce electron transitions between spin states, particularly when the microwave energy matches the energy difference between states. The resulting spectrum, characterized by discrete lines at specific magnetic fields, reflects the total number of defects in the EPR-active paramagnetic charge state. EPR can quantify small defect concentrations (as low as 1 ppb) including unpaired electrons or holes, as well as various defects and extrinsic impurities like donors and acceptors [339–351]. EPR spectroscopy stands out among characterization techniques like electron microscopy and X-ray absorption for its unique capability to analyze bulk materials rather than just surfaces [284].

EPR spectroscopy is particularly effective in characterizing paramagnetic atoms in dilute concentrations within crystalline hosts. It is especially adept at detecting Kramers ions, such as $Fe^{3+}$, $Mn^{2+}$, and $Mn^{4+}$, which have an odd number of unpaired spins. These ions are more easily observable using standard X-band EPR equipment, unlike non-Kramers ions like $Mn^{3+}$, which are often EPR silent [352–355]. It sheds light on the local environments of these ions, especially the nearest neighbor oxygen ligands. This capability is crucial in investigating oxygen vacancies [349]. While EPR can't directly identify oxygen vacancies, especially in dilute concentrations, EPR provides straightforward analysis of their interaction with paramagnetic defects [356].

EPR spectroscopy's historical significance in defect characterization is highlighted by its application in studies on Fe-doped STO in the late 1950s [357]. Pioneering work of Müller first derived and measured the electron spin Hamiltonian for the $Fe'_{Ti}$ defect [357], Subsequently, extensive EPR studies on various defect complexes have greatly advanced our understanding of defect dipoles [262,339–342,347,349–351,358–363]. A notable example is the work by Warren et al. on undoped BTO single crystals [364]. This study was groundbreaking in its detection of $Fe^{3+}$ centers, which were presumed to arise from acceptor-type impurities. Furthermore, EPR can monitor both the reorientation of defect dipoles via short range motion of oxygen vacancies, and diffusion of oxygen vacancies over long distances under an external field. The first phenomenon is instrumental in creating internal fields, while the latter plays a crucial role in enhancing ionic conductivity in perovskite oxides. The interaction between $(Mn''_{Ti} - V_O^{\bullet\bullet})$ defect dipoles, domain structure and an external field was studied in Mn doped single crystal BTO, as shown in Figure 13 [262,345,365]. The EPR spectrum of unpoled sample comprises two sets of hyperfine multiplets from $Mn^{2+}$, corresponding to two magnetically equivalent $Mn^{2+}$ centers with different orientations relative to the external magnetic field. Upon poling, the sample transitioned into a single-domain state. Consequently, the EPR spectrum was characterized by a single Mn center, which indicates that $(Mn''_{Ti} - V_O^{\bullet\bullet})$ defect dipoles were oriented in the same direction.

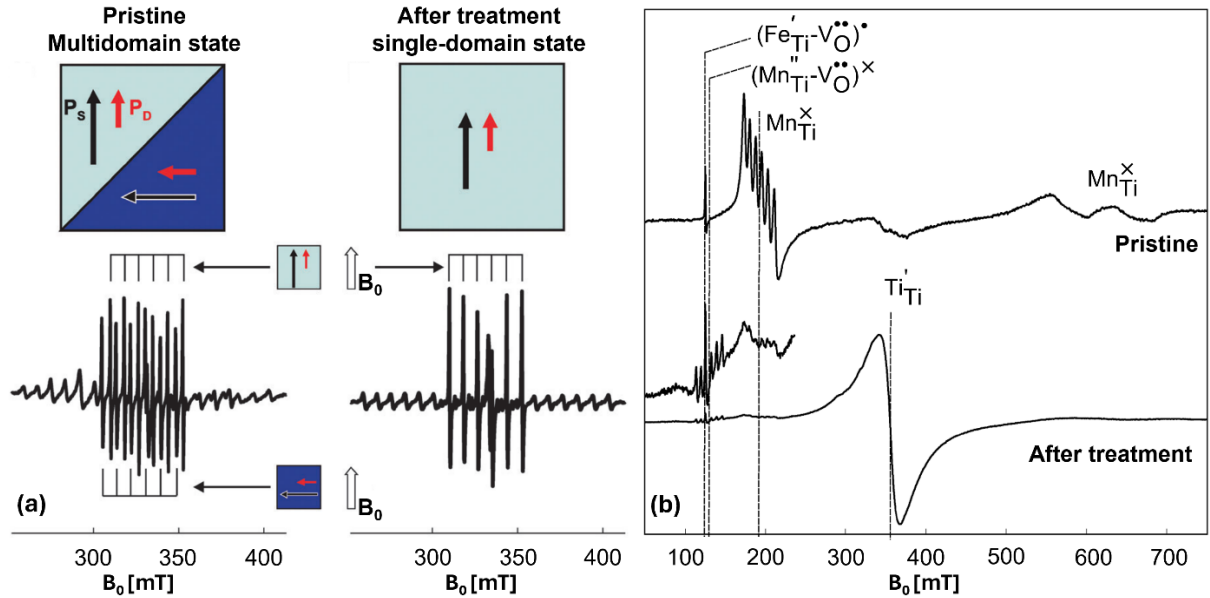

Figure 13. (a) EPR spectra of $(Mn''_{Ti} - V_O^{\bullet\bullet})$ defect dipole centers in a BTO single crystal in a multidomain (left), and a poled single domain state (right), (b) EPR spectra of Mn doped PbTiO$_3$ in pristine state (top) and after heating in reducing atmosphere (bottom). Reproduced with permission from ref. [262].

The interaction between multivalent functional centers and electronic or ionic charge carriers was analyzed in Mn doped PbTiO$_3$ ceramics exposed to heat treatment under varying oxygen partial pressure conditions. The EPR spectrum of pristine Mn-doped PbTiO$_3$ consisted of only $Mn^X_{Ti}$ centers. After annealing under a reducing atmosphere, the Mn$^{4+}$ signal largely vanished, with a concurrent emergence of Ti$^{3+}$ and $(Mn''_{Ti} - V_O^{\bullet\bullet})$ defect dipole signals due to oxygen loss Each oxygen vacancy introduces two electrons that can be trapped by Ti$^{4+}$ ions to generate Ti$^{3+}$. The presence of $(Mn''_{Ti} - V_O^{\bullet\bullet})$ defect dipoles in reduced samples indicates that the oxygen vacancies are trapped by $Mn''_{Ti}$ ions, significantly affecting ionic conductivity [346,366].

The ability of EPR to discern the alignment of defect dipoles offered profound insights into the dynamics of these dipoles under different conditions, including those associated with electrical degradation. As a result, it has been possible to directly enhance material performance and reliability [197,269,367,368].

### 4.1.6 Positron Annihilation Lifetime Spectroscopy (PALS)

Positron Annihilation Lifetime Spectroscopy (PALS) is an effective method for detecting and quantifying vacancy environments, ranging from monovacancies to large vacancy clusters [271–274]. In PALS experiments, a $^{22}$Na radioactive source is typically used to produce positrons. $^{22}$Na undergoes β$^+$ decay, resulting in the emission of a positron and a neutrino. This leads to the formation of an excited state of $^{22}$Ne, which rapidly decays by emitting a 1.2745 MeV γ photon. This photon is used to initiate the timing for the positron's lifetime measurement. The experiment concludes upon detection of one of the dual 511 keV gamma photons, which are emitted when the positron annihilates[369]. In PALS, positrons are introduced into a material that can interact with electrons and form a quasiatom, known as positronium, and eventually annihilate with these electrons, emitting gamma rays in the process. The key parameter measured is the lifespan of the positron from its introduction to annihilation, which reflects the local electron density at the annihilation site [274]. Trapping of positrons at defects results in an increased positron lifetime and a narrower momentum distribution width of the annihilating positron-electron pair, known as Doppler broadening [370]. The broadened spectrum's shape is analyzed using two parameters: S and W. The S parameter represents the proportion of counts in the central part of the peak, while W measures the counts in the wing areas on either side of the peak, being particularly sensitive to core electrons that are highly localized [370]. An increase in the S parameter generally signals a rise in open-volume defects like voids or vacancies within the analysis volume. Conversely, the W parameter is more indicative of the types of atoms surrounding the annihilation site, thus giving insights into the defect's local atomic environment [274,275].

PALS has exceptional sensitivity to atomic-scale defects and microstructural changes. Its non-destructive nature preserves the sample integrity during analysis. The technique's versatility is enhanced by slow positron technology, which facilitates adjustable energy levels for probing defects or inhomogeneities in structure. However, PALS faces limitations in detecting positively charged defects and determining defect charge states. To address this, PALS is often used alongside complementary methods like XPS [276,277].

Rutkowski et al. [370] employed PALS, Depth-Resolved CL Spectroscopy (DRCLS), and Surface Photovoltage Spectroscopy (SPS) to distinguish between defects in (Ba,Sr)TiO$_3$ epilayers from those in interface regions near the STO substrate. The 2.95 eV peak, frequently observed in STO, has been linked to an oxygen vacancy or a defect complex associated with an oxygen vacancy is 2.95 eV above the valence band. DRCLS showed that this 2.95 eV luminescence intensifies near the heterojunction and becomes more pronounced as $P_{O_2}$ decreases during growth [370]. For defects in the (Ba,Sr)TiO$_3$ epilayers, DRCLS and SPS revealed an inverse relationship between barium vacancies and oxygen vacancies. PALS was used to explore oxygen vacancy-related defects; by examining changes in the S and W parameters, Rutkowski et al. [370] identified variations in defect concentration at different depths. They observed a significant increase in these parameters at positron implantation energies corresponding to the (Ba,Sr)TiO$_3$/STO interface depth. These observations were explained by the presence of oxygen vacancies clusters, including oxygen divacancies and complexes involving metal vacancies and oxygen vacancies, forming near the interface.

### 4.2 Detection of Trap Levels: Probing Defect Chemistry for Degradation Analysis

#### 4.2.1 Charge-Based Deep Level Transient Spectroscopy (Q-DLTS)

Deep Level Transient Spectroscopy (DLTS), established by Lang in 1974 [371], remains an essential technique for the characterization of trap levels in semiconductor devices, including p-n and p-i-n junctions and Schottky diodes [274,372–378]. This method excels in identifying and analyzing electrically-active defects through capacitance transients as a function of temperature. DLTS distinguishes between minority and majority carrier traps, providing insight into defect densities, their emission and capture dynamics. There are two principal approaches to DLTS: capacitive DLTS, which focuses on changes in the thickness of the depletion layer, and charge-based DLTS (Q-DLTS), which offers increased sensitivity to bulk and interfacial traps [379]. C-DLTS is not applicable to ferroelectric materials due to non-linear dependence of the permittivity with respect to the electric field in the depletion regime. Thus, Q-DLTS has become the method of choice for investigating defect levels and local electronic states in dielectric materials under transient electrical pulses [380–382]. This method is based upon measuring space charges from traps as a transient charge. This technique is very similar to Sawyer-Tower methods used in the characterization of switchable polarization in ferroelectric materials [379]. The emission rate €, or the corresponding time constant (τ=1/e), for charge release, whether for holes or electrons, at a given temperature T, follows an exponential decay characterized by [382,383]:

$$e_{p,n} = \tau_{p,n}^{-1} = \sigma V N_c T^2 exp\left(-\frac{E_a}{k_B T}\right) \quad (21)$$

where σ represents the capture cross section, V and $N_C$ are the thermal velocity and the density of states, respectively. $E_a$ is the trap activation energy which quantifies the thermal activation process of carrier emission. The relaxation of transient charge during discharge for a constant rate window defined by times $t_1$ and $t_2$ at each temperature, as illustrated in Figure 14.a, is expressed as:

$$\Delta Q = Q(t_1) - Q(t_2) = Q_0[exp(-e_{p,n}t_1) - exp(-e_{p,n}t_2)] \quad (22)$$

From these equations, the trap parameters σ and $E_a$ can be determined by recording the Q-

DLTS as a function of temperature. Each Q-DLTS peak is fitted using a Gaussian distribution function. Alignment of the experimental and theoretical peak central positions, together with examining different rate windows, allows estimation of parameters such as $E_a$ and $\sigma$ within the context of an Arrhenius equation [371,382]. However, despite its apparent simplicity, this method demands multiple data points (usually 5–7 minimum) to ensure accurate extraction of parameters like activation energy. Achieving this requires scanning a broad range of rate-windows and temperatures, necessitating DLTS electronics that can measure nanoamp-level discharge current and precise cryogenic temperature control apparatus (0.05-0.2 eV activation energies can be determined at 150-250 K and 0.2-2 eV can be deduced at temperatures between 300-800 K [269]). Finally, the trap density $N_t$ is determined by measuring the value of charge maximum ($\Delta Q_m$) and by using the following equation [8]:

$$N_t = \frac{4\Delta Q_m}{qSd} \tag{23}$$

where q is the electronic charge ($1.602 \times 10^{-19}\ C$); S and d correspond to the active surface area and the thickness of the sample, respectively.

The changes in trap center characteristics and their densities as a function of dopant species and levels in PZT films were investigated using Q-DLTS. Figure 14.b shows the Q-DLTS signals of 2 mol% Mn doped PZT films (with a Zr/Ti ratio of 52/48) from 100-650 K. Three pronounced peaks were observed, where the low temperature peak correlating to a small activation energy of 0.05 eV is likely indicative of hopping conduction. The intermediate temperature trap level, exhibiting an activation energy of 0.26±0.05 eV, has been ascribed to hole hopping between $Pb^{2+}$ and $Pb^{3+}$ lattice sites as suggested by Robertson et al. [203]. The high temperature Q-DLTS peak, with an activation energy of 1.9±0.2 eV, corresponds to the $Mn^{2+/3+}$ impurity states within the band structure of PZT. The $Mn^{2+/3+}$ trap density progressively rises on increasing Mn doping level in the PZT film. In PZT films doped with up to 6 mol% Nb, the predominant impurity sites contributing to enhanced electronic conductivity are the trap levels involving $Pb^{2+}/Pb^{3+}$ transitions. Beyond a niobium concentration of 6 mol%, the mechanism governing electronic conduction shifts to predominantly hole migration via lead vacancies, $V_{Pb}''$ [250,261,269].

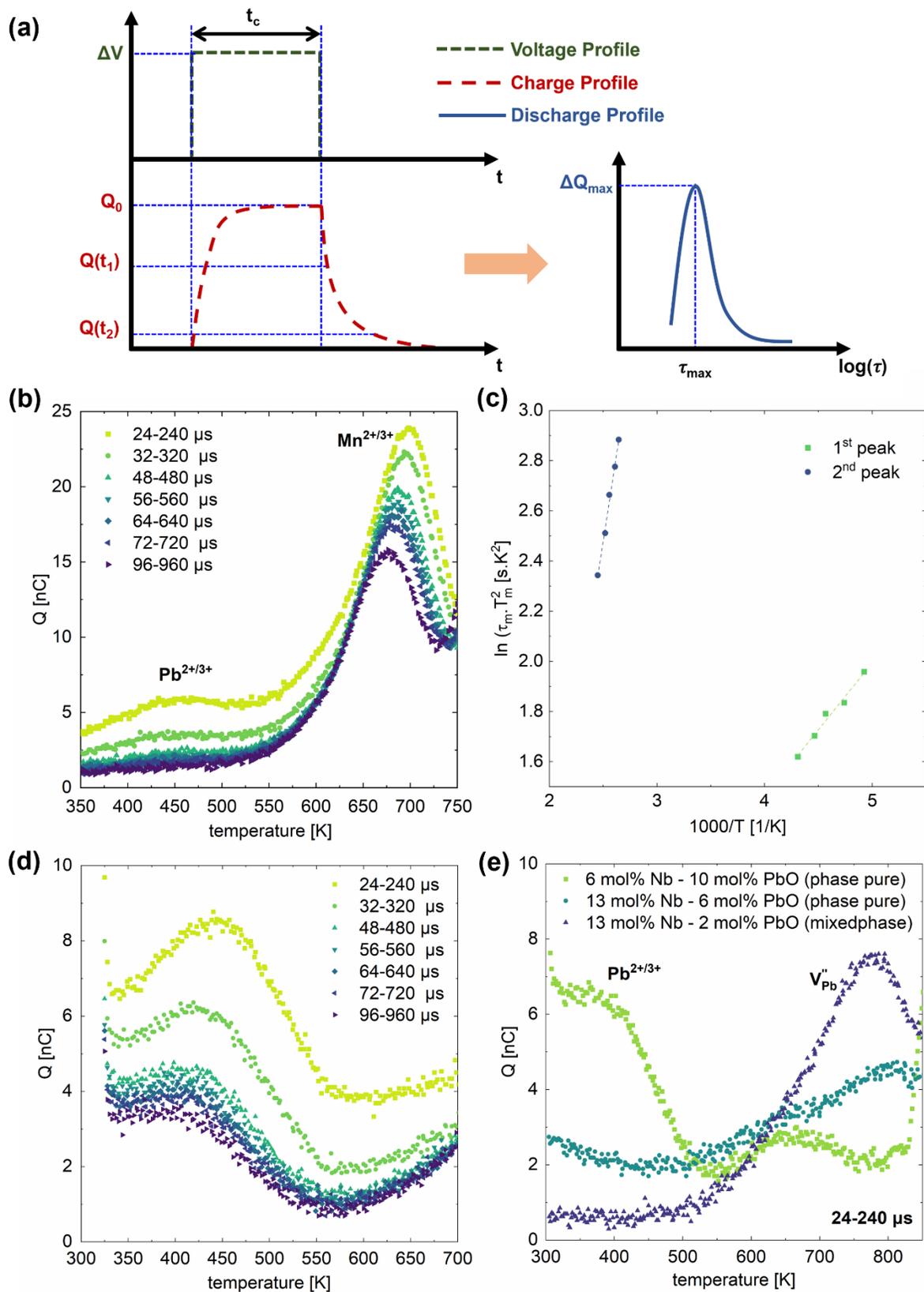

Figure 14. (a) The implementation of the rate window concept using a double boxcar integrator, where the output represents the average difference in charge amplitudes at sampling

times t$_1$ and t$_2$, (b) Q-DLTS analysis of 2 mol% Mn-doped PZT films conducted across a temperature range of 300 to 750 K. The DLTS signal was measured at seven distinct rate windows, aimed at resolving emissions from various traps, (d) Q-DLTS results for 2 mol% Nb-doped PZT films at different window rates, (e) Q-DLTS data for both phase-pure 6 and 13 mol% Nb-doped PZT films (in green), as well as mixed-phase 13 mol% Nb-doped PZT films (in blue) (redrawn from data in ref. [269]).

DLTS as a technique needs to evolve in order to address non-Arrhenius processes. Such processes become evident when defect signatures persist across a broad rate-temperature range. This can be due to a temperature-dependent activation energy or capture cross-section [384]. In response, techniques like temperature-rate duality relationships have been introduced, enabling independent extraction of activation energies and attempt-to-escape frequencies, bypassing the constraints of conventional Arrhenius plotting [385]. Additionally, Laplace-transform DLTS (LDLTS) offers significantly higher energy resolution compared to standard DLTS methods, primarily attributed to its superior signal-to-noise ratio [274]. However, optimal performance of LDLTS depends on having a significant concentration of the defect under investigation within the sample [386]. LDLTS excels in differentiating between defects with closely related carrier emission rates, a task unachievable with traditional DLTS, due to reduction of instrumental broadening. Furthermore, when used in tandem with uniaxial stress, LDLTS provides valuable insights into defect symmetry. This capability is exemplified in studies of oxygen - vacancy pairs in silicon, where LDLTS can be effectively applied to extremely thin semiconductor regions, a challenging endeavour for other characterization methods [387]. Such comprehensive insight into defect characterization is crucial in understanding and mitigating electrical degradation in devices.

### 4.2.2 Cathodoluminescence and Photoluminescence Spectroscopy

Cathodoluminescence (CL) is a technique that involves light emission from a sample when excited by an electron beam. CL shares similarities with both electron probe microanalysis (EPMA) and photoluminescence. Its advantage lies in its imaging capabilities, where the electron beam scans across a sample, and the emitted light is detected. The key difference between EPMA and CL is the source of emitted photons: EPMA involves transitions between inner core energy levels, whereas CL emissions result from transitions between conduction and valence bands [274]. Quantitative interpretation of CL images can be challenging due to spatially dependent factors like local reflectance or surface morphology. Moreover. the intensity of light emission in CL can be influenced by several factors, including local temperature variations, recombination centers associated with impurities and defects, and doping densities. Additionally, the presence of an electric field affects the CL emission, adding another layer of complexity to the interpretation of these images [274]. Basic CL involves room temperature analysis, but cooling and spectral resolution enhance its effectiveness. Cooling reduces thermal line broadening, improving signal-to-noise ratio, while spectral resolution aids in impurity detection. CL's resolution depends on the electron beam diameter, range, and minority carrier diffusion length [274]. Time-resolved CL helps measure lifetimes. Quantitative interpretation is complicated by factors like local reflectance variations and surface morphology.

Recent studies have even suggested the presence of an antisite Ti defect ($Ti''_{Sr}$), where a Ti atom replaces a Sr atom in STO thin films prepared under metastable processing conditions [108,109]. This was further investigated by Lee et al. [388] using a combination of methods, including cathodoluminescence spectroscopy [90], as shown in Figure 15. The CL measurements uncovered sub-band-gap luminescence, with spectral fine structures that closely matched theoretical predictions of optical transitions from point defects like $Ti''_{Sr}$. Wang et al. [301] also used CL spectrometry to study the influence of dislocations on oxygen vacancies in undoped STO single crystals. They identified a prominent luminescence peak ~2.8 eV, indicative of oxygen vacancies. These authors also noted enhanced luminescence at specific dislocations, inferring that these are local sites driving accumulation of oxygen vacancies. Furthermore, on annealing in either oxidizing or reducing conditions, the mobility of oxygen ions at dislocation cores mirrors that within the bulk STO, thereby suggesting that dislocations prompt accumulation of oxygen vacancies but without significantly altering their diffusion in STO.

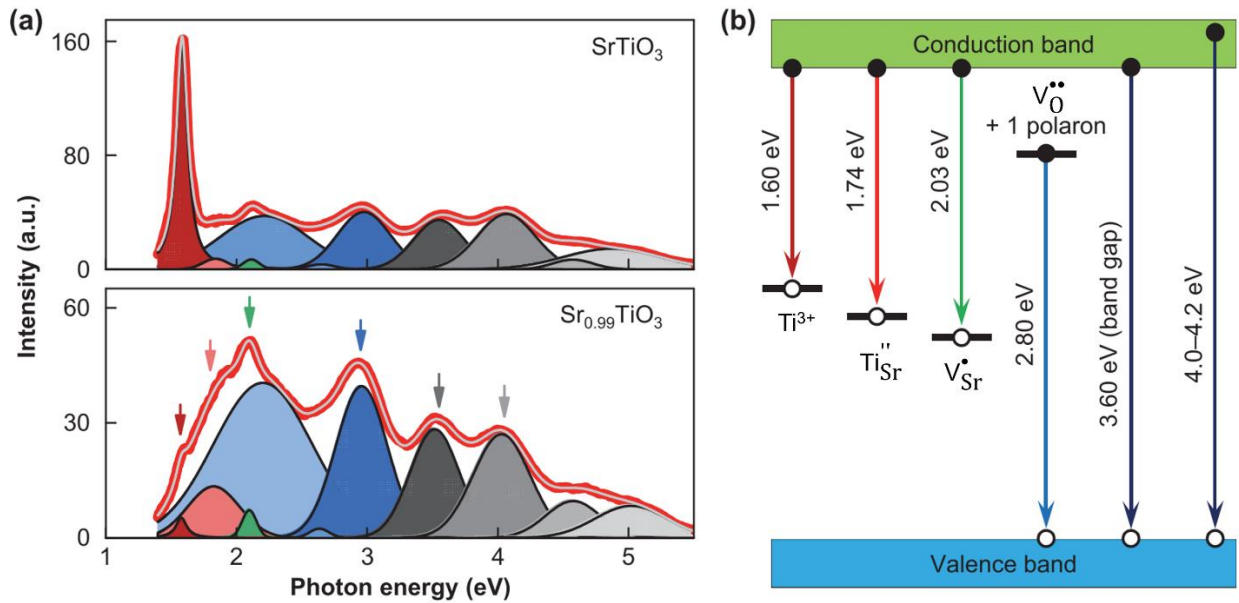

Figure 15. (a) Comparison of cathodoluminescence spectra of 24-unit-cell-thick homoepitaxial STO and Sr$_{0.99}$TiO$_3$ films. (b) Schematics of luminescence processes. Reproduced with permission from ref. [388].

Chen et al. [389] similarly employed CL spectroscopy to focus on the behavior of oxygen vacancies during the degradation of BTO-based ceramics. As shown in Figure 16, the CL spectrum of BTO-based sample revealed five emission peaks. The first peak at 425 nm (Peak I) aligns with BTO's energy bandgap which can range from 2.9–3.4 eV depending on Ba/Ti ratio, the dopants and their concentrations, and the concentration of oxygen vacancies [86,390–393]. The 490 nm emission (Peak II) is attributed to recombination of electrons trapped by oxygen vacancies, with valence band holes. Peak III, emitting at 550 nm, results from conduction band electrons recombining with holes trapped by barium vacancies. The 620 nm emission (Peak IV) originates from electrons trapped by oxygen vacancies recombining with holes trapped by barium vacancies. Finally, the 750 nm emission (Peak V) is due to conduction band electrons recombining with holes trapped by doubly ionized barium vacancies. In acceptor-doped perovskite ceramics, doubly ionized oxygen vacancies are common. However, their transitions to the valence band are improbable, making it difficult to detect emissions from these vacancies in the valence band using CL in BTO. Their observations also indicated that, during degradation, oxygen vacancies in grains without a core-shell structure quickly moved past grain boundaries and migrated towards cathode-adjacent grains. Conversely, in grains with a core-shell structure, oxygen vacancy migration was largely confined to the grain core in the early to middle stages of degradation, effectively slowing down the degradation process.

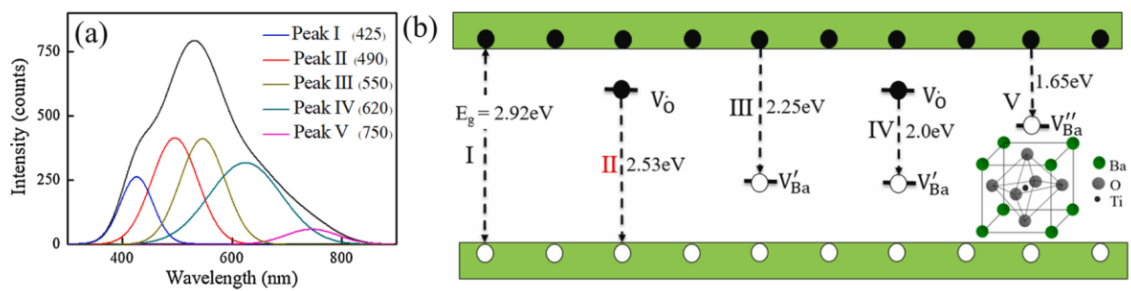

Figure 16. (a) Cathodoluminescence (CL) spectrum detected from the grain in the BTO-based ceramics fitted with five Gaussian peaks and (b) Possible mechanisms for luminescence emission. Reproduced with permission from ref. [389].

DRCLS is used to analyze the resistance degradation mechanisms in materials [370,394–397]. This method offers insight into the nanoscale distribution of defects and their electronic properties, particularly under applied electric fields or at elevated temperatures. The process begins with electron irradiation of the solid, leading to a cascade of secondary electrons. These electrons undergo various energy loss mechanisms, including Auger electron excitation, secondary electron generation, backscattering, and X-ray generation [398–402]. The primary mode of energy loss changes depending on the energy of the incident electron beam: X-ray generation becomes the dominant process at energies exceeding a few tens of keV, whereas plasmon generation becomes more notable at energies below 100 eV. Towards the end of the energy cascade, the electrons attain energies only adequate for impact ionization, resulting in the creation of electron-hole pairs [394]. The depth of optical excitation by the incident beam is determined by the distribution of low-energy electrons, typically spanning tens to hundreds of nm [394,403,404].

Gao et al. [394] utilized DRCLS focusing on oxygen vacancy-related defect complex behavior pre and post electric field-induced degradation in undoped and Fe-doped STO. DRCLS provided insights into the spatial distribution and energy levels associated with specific defect complexes in three dimensions on a nanometer scale. It also directly detected the electromigration of oxygen vacancies from anode to cathode across laterally separated electrodes in undoped STO under applied electric fields, leading to local carrier compensation and enhanced electrical conductivity. Figure 17 detailed the energy levels within the bandgap of STO in relation to degradation processes. Gao et al. [394] observed a 0.1 eV spectral shift in optical transition energies (from 2.55 eV to 2.65 eV) and a 0.06 eV decrease in the free carrier activation energy, which corresponds to changes in oxygen vacancy defect configuration during electromigration, leading to electrical degradation. Despite the proximity of these levels to $Fe^{3+}$ states in Fe-doped

STO, they do not exactly coincide, suggesting distinct defect behaviors. Furthermore, XPS measurements reveal a range of Fermi level positions in (Ba,Sr)TiO$_3$, consistently above midgap, irrespective of doping type or synthesis route. These findings highlight the complex interplay of defect states, particularly oxygen vacancies, with the electronic structure of STO, affecting its degradation under stress conditions.

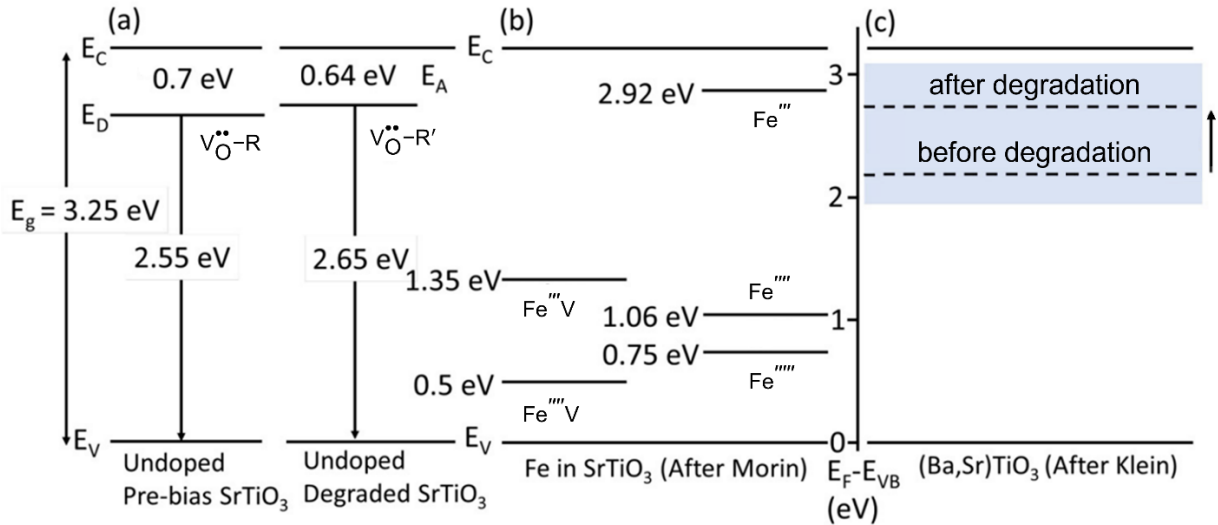

Figure 17. Schematic representations of (a) Defect transitions in STO measured by CL before and after electrical degradation, showing energy level shifts and $V_O^{\bullet\bullet}$-R complex movement under degradation; (b) Primary defect states of Fe complexes in STO; (c) Experimentally obtained Fermi level positions in doped (Ba,Sr)TiO$_3$ and STO via XPS; that revealed the shift in Fermi level (dashed lines) at the cathode from 2.2 to 2.7 eV, as indicated by arrow. Reproduced with permission from ref. [394].

Photoluminescence spectroscopy (PL) is a non-destructive technique particularly effective for detecting shallow or deep-level impurity states that undergo radiative recombination. It operates by generating electron-hole pairs with incident radiation, followed by their radiative recombination. Identifying impurities using PL is straightforward, but measuring their density can be challenging due to recombination centers that affect the PL signal intensity. Recent advances in PL, such as simultaneous intrinsic and extrinsic analysis, have improved impurity density estimation [274,405,406].

In materials like BTO and STO, PL is observed due to electron-hole recombination in intermediate states, often linked to distorted clusters with oxygen vacancies [407]. As electrons within these intermediate electronic states combine with holes derived from the valence band, this process induces radiative recombination, giving rise to the detection of photoluminescent emission lines [407–413]. For example, pure BTO exhibits a broad visible luminescent band at low temperatures when excited by radiation exceeding its energy band gap. In nanocrystalline BTO,

PL has also been observed at room temperature, even with excitation energies below the energy band gap. This PL behavior in nanocrystalline BTO is attributed to the size of the particles and the presence of oxygen vacancies [405,414–416].

### 4.3 Detecting and Analyzing Degradation and Breakdown

#### 4.3.1 Electrochemical Impedance Spectroscopy

Electrochemical Impedance Spectroscopy (EIS) has evolved significantly since its inception in the late 19$^{th}$ century. Advances in computer-control and digital electronics facilitated rapid measurements, extensive data processing, and sophisticated analysis, significantly amplifying its utilization. While comprehensive reviews of EIS can be found in literature [417–427], this section focuses on the application of EIS in probing defect chemistry and gaining insights into electrical degradation assessment.

EIS is a powerful, non-destructive technique for unravelling materials' electrical responses and their connections to microstructural attributes. This technique entails applying a sinusoidal current in galvanostatic mode, or a sinusoidal voltage in potentiostatic mode, across the material. It then measures the amplitude and phase shift of the resulting output in voltage or current, respectively. Maintaining linear proportionality between the input and output signals, with the input potential energy around the order of $k_bT$, remains crucial [426]. For steady-state EIS measurements that take several minutes to hours, it is important that the applied field not change the sample during data collection [425]. Unlike DC resistance measurements, which provide overall resistance values but cannot differentiate between various microstructural contributions, EIS probes different physical and microstructural aspects of the material, enabling a more nuanced and detailed understanding of its electrical properties and behavior. The complex impedance is given by:

$$Z^*(\omega) = Z'(\omega) - Z''(\omega) \qquad (24)$$

where Z', Z'', ω are the resistance, the reactance, and the angular frequency, respectively. The Debye model assumes a singular relaxation time within the system and describes the frequency dependence of impedance with a Gaussian distribution of relaxation times. This is represented as [39]:

$$Z^*(\omega) = R_\infty + \frac{R_0 - R_\infty}{1 + i\omega\tau_0} \qquad (25)$$

where $R_0$ and $R_\infty$ are the low and high frequency resistances, respectively. $\tau_0$ is the characteristic relaxation time ($\omega\tau_0=1$), and the parameter α is a measure of the sharpness of the distribution of the relaxation times, which can be expressed for a Gaussian distribution as [428]:

$$F(s) = \frac{1}{2\pi}\frac{\sin\alpha\pi}{\cosh(1-\alpha)s - \cos\alpha\pi} \tag{26}$$

where $s = lin(\tau/\tau_p)$.

EIS data are commonly shown in either a Nyquist plot representing the imaginary part against the real part of the impedance, or a Bode plot mapping magnitude and phase or real and imaginary parts of the response as functions of frequency, as shown in Figure 18. These plots aid in parameterizing the admittance (Y), permittivity (ε), modulus (M), and capacitance (C). The interrelationships between these parameters are detailed in Table 1. Notably, EIS measurements can be made as a function of temperature, enabling *in-situ* measurements, providing real-time insights unattainable through post-mortem characterization methods.

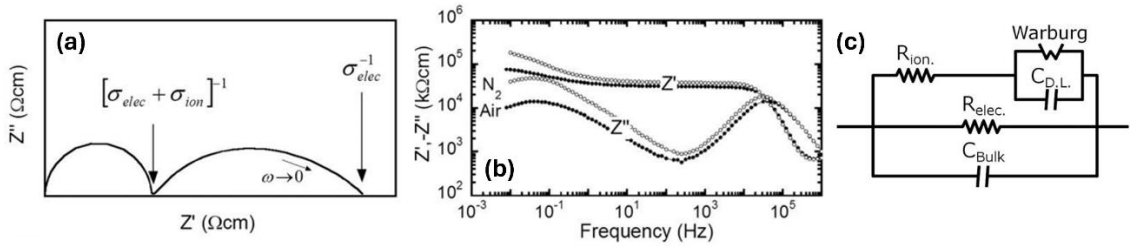

Figure 18. Schematic of impedance spectrum of system with mixed ionic-electronic conduction: (a) Nyquist plot, (b) Bode plot, and (c) an effective equivalent circuit used to model this system, where $R_{elec.}$ and $R_{ion}$ are the electronic and ionic resistance, respectively, $C_{Bulk}$ and $C_{D.L.}$ are the dielectric and double layer capacitance respectively. Reproduced with permission from ref. [429].

Table 1. Relations between dielectric parameters in EIS, where $\tilde{C}$ is the complex capacitance and $C_0$ is the capacitance of the empty cell ($C_0 = \varepsilon_0 A/d$, where A, d, and $\varepsilon_0$ are cell area, thickness, and the permittivity of free space, respectively.) [426].

|  | unit | C | Z | M | Y | ε |
|---|---|---|---|---|---|---|
| Capacitance (C) | F | $\tilde{C}$ | $(i\omega Z)^{-1}$ | $C_0 M^{-1}$ | $(i\omega)^{-1} Y$ | $C_0 \varepsilon$ |
| Impedance (Z) | Ω | $(i\omega\tilde{C})^{-1}$ | Z | $(i\omega C_0)^{-1} M$ | $Y^{-1}$ | $(i\omega C_0 Z)^{-1}$ |
| Dielectric modulus (M) | m/F | $C_0 \tilde{C}^{-1}$ | $i\omega C_0 Z$ | M | $i\omega C_0 Y^{-1}$ | $\varepsilon^{-1}$ |
| Admittance (Y) | S | $i\omega\tilde{C}$ | $Z^{-1}$ | $i\omega C_0 M^{-1}$ | Y | $i\omega C_0 \varepsilon$ |
| Permittivity (ε) | F/m | $C_0 \tilde{C}$ | $(i\omega C_0 Z)^{-1}$ | $M^{-1}$ | $(i\omega C_0)^{-1} Y$ | ε |

Understanding the kinetic processes in electroceramics or electrochemical systems relies on establishing a suitable analog model using equivalent circuits. These circuits, composed of series and/or parallel configurations of resistors and capacitors, provide valuable insights into defect chemistry and material behavior by linking impedance data to physical parameters [47,426]. As Abram et al. [352] pointed out, these formalisms involve varying weighting factors on diverse combinations of circuit components. Consequently, proper modeling provides a strong foundation for establishing the physical model of the systems.

The Debye theory of dielectric loss connects a material's electrical behavior, such as ion diffusion across a grain boundary, to the frequency variation seen in Bode plots (Figure 18.b). Each peak in the impedance versus frequency plot correlates to a semicircle in the complex plane (Figure 18.a). Distinct relaxation time scales are evident as individual semicircles or peaks in the impedance spectra, offering insights into the number of relaxation processes within the spectrum. Conversely, overlapping semicircles suggest similar relaxation time scales, necessitating data modeling for differentiation. Where there are common distributions of time constants within inhomogeneous systems, the arcs below the real plane of the Nyquist plot can be suppressed [39]. Constant Phase Elements (CPEs) can be employed to simulate time constant distributions in various circumstances. The impedance of a CPE is defined by two parameters, T and P:

$$Z_{CPE} = \frac{1}{T(i\omega)^P} \qquad (27)$$

where P signifies a dispersion parameter that determines the physical meaning of T. When P=1, the CPE emulates an ideal capacitor with T representing the capacitance [426]. In contrast, when P=0, the CPE transforms into a real resistor with T being equivalent to the inverse of the resistance (1/R). At P = 0.5, the CPE mirrors a Warburg element, and T corresponds to the reciprocal of the Warburg coefficient ($1/A_w$), determined by the AC diffusion coefficient. The Warburg element is characterized by a linear increase in Z values at low frequencies, and so has a phase angle ($\phi$) of 45° on the Nyquist plot. The CPE is employed to characterize complex relaxation phenomena characterized by multiple time scales, such as time-dependent or dispersive processes like charge transfer occurring on nonuniform electrode surfaces or carrier transport within amorphous semiconductors. In cases involving relaxation time distributions (where 0.5 < P < 1), the conventional ideal capacitor is substituted with a CPE element, with T=C, and P>0.5, integrated into the circuit. [430]. When a sample exhibits multiple relaxation times, like grain and grain boundary reactions, each is characterized using a parallel combination of a resistor and CPE, enabling separate analysis [13,58,161,166,201,429,431,432]. In ceramic grains, C is typically around $10^{-12}$ F, while grain boundaries exhibit notably higher C values from $10^{-11}$ to $10^{-8}$ F, and large sample-electrode interfaces yield even larger C values from $10^{-7}$ to $10^{-4}$ F [433].

EIS is adept at differentiating between contributions from grains, grain boundaries, and electrode/dielectric interfaces to the overall conductivity in polycrystalline bulk and thin film materials, particularly those exhibiting ionic or mixed ionic-electronic conduction [341,342]. Maier et al. [343] outlined common equivalent circuits, Figure 19, that can be used to describe mixed ionic/electronic conduction in capacitive ceramics. For instance, in donor doped BTO Positive Temperature Coefficient Resistors (PTCRs) and BME capacitors, *n*-type carriers predominate [434]. In such systems, interfaces significantly influence the impedance response at lower temperatures, with Maxwell-Wagner relaxation arising due to inhomogeneous electrical conductivity. This relaxation is often characterized by grain-grain boundary polarization, especially in doped BTO ceramics. Likewise, for p-type STO, the impedance response is described by the accumulation and depletion of mobile charge carriers at interfaces between highly conductive grains and poorly conductive grain boundaries. These grain boundaries act as double Schottky barriers, leading to space charge polarization [435]. Assuming the grain boundary resistance ($R_{GB}$) and capacitance ($C_{GB}$) are significantly higher than the grain resistance ($R_G$) and capacitance ($C_G$), the current decay ($\tau_r$) can be expressed by the charging of the grain boundary capacitance [47]:

$$\tau_r = R_G * C_{GB} \tag{28}$$

Therefore, the most resistive component controls the overall conductivity. Grains with core-shell structures can have different resistivities and thereby give two further relaxations within the impedance response [436]. At higher temperatures, dielectrics can also be treated with mixed conduction, provided there are ionically blocking metallic electrodes and non-blocking electronic species in the impedance measurement. If these conditions are not met, large conductivity changes and gas evolution occur, which can make interpretation difficult. Therefore, equivalent circuit models must avoid departures from defined boundary conditions. Figure 18.c represents an effective equivalent circuit used to model a system with mixed ionic-electronic conduction and ion blocking electrodes [198]. The ionic branch in the equivalent circuit is conductive at high frequencies and sufficiently high temperatures. The total resistance is equal to the combination of electronic resistance, $R_{elec.}$, ionic resistance $R_{ion}$, and the conductivity calculated from the geometric factors. The dielectric capacitance, $C_{Bulk}$, generates the high-frequency semicircle in the $Z^*(\omega)$ plot. At lower frequencies, the ionic carriers have time to accumulate at the electrodes, and the impedance approaches $R_{elec}$ in the DC limit. The low-frequency relaxation is generated from the Warburg and/or double layer capacitance, $C_{D.L.}$, elements that are in series with the ionic resistance/conductance. This form of data is especially useful in mixed conductors as the respective conductivities are easily extracted from the intercept points on the real axis. These are used to quantify the transference number, the conductivity ratio between the respective components of conductivity. Warburg impedance provides diffusion-limited behavior, while double layer capacitance relates to the ionic space charge polarization at the electrode [144,198,428].

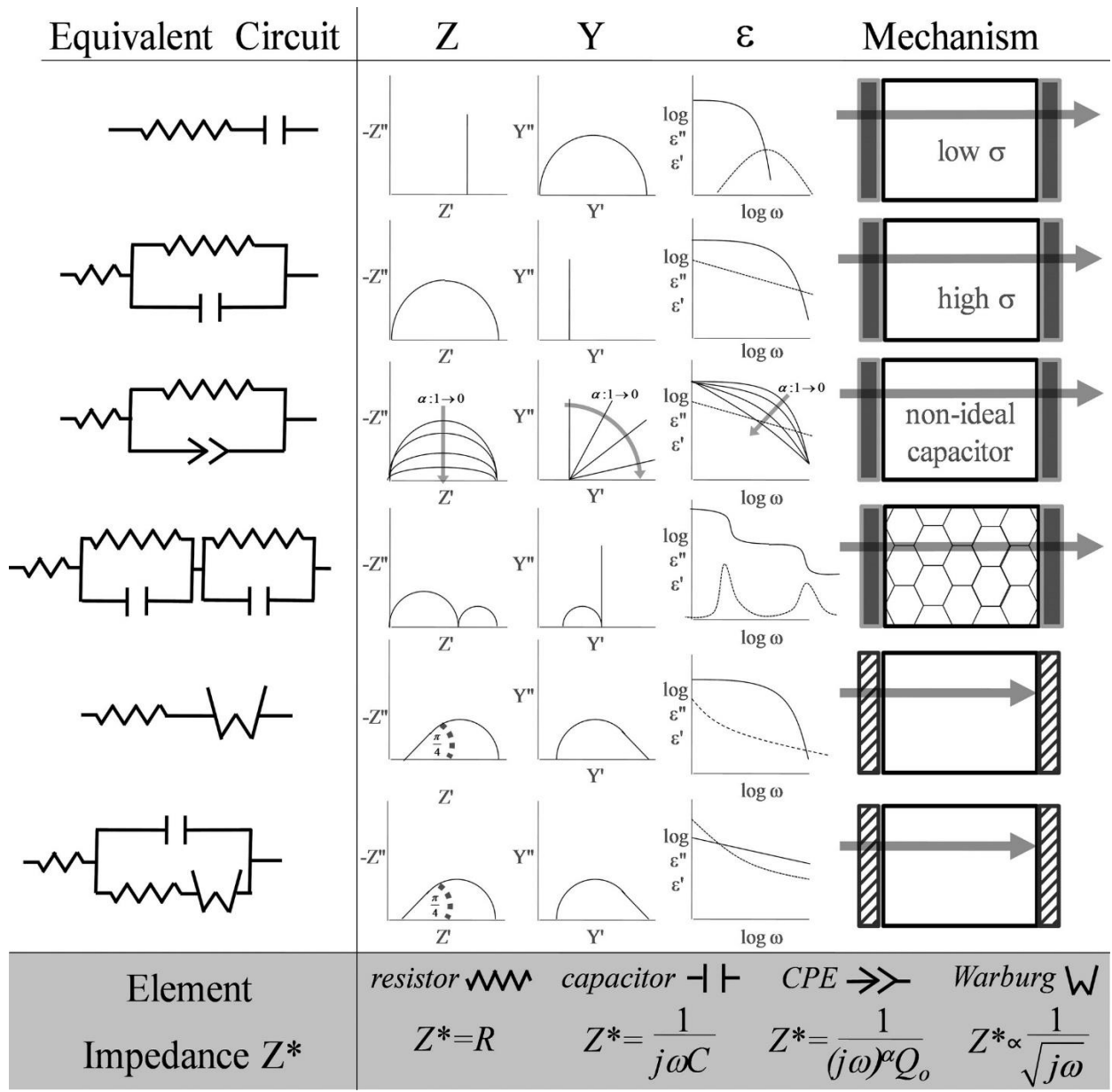

Figure 19. Complex impedance responses illustrate different electrical circuits commonly employed to characterize mixed ionic/electronic conduction in capacitive ceramics. The right column provides schematics depicting the respective dielectric system responsible for each response. Reproduced with permission from ref. [434].

For many capacitive ceramics, degradation reduces the electrical resistivity of the electrode interface, grain boundaries, and grains at different rates [81,437,438]. Chazono et al. [180] studied the electrical degradation of BTO-based MLCCs and utilized a 4-RC electrical equivalent circuit to fit the impedance data, as shown in Figure 20. Each RC circuit corresponds to a different microstructure component of a BTO-based MLCC. They found that $R_1C_1$, $R_2C_2$, $R_3C_3$, $R_4C_4$ are correlated to shell region, grain boundaries, and ceramic/internal electrode interface, and core region respectively. They observed that the capacitances of both the core and shell regions followed the Curie-Weiss law; the inverse square of the capacitance for the grain boundary (GB) region exhibited a linear increase with the applied DC voltage. In contrast, the capacitance for the ceramic/internal electrode interface region increased with rising DC voltage, suggesting different characteristics from those of the GB region. Their study revealed that leakage current behavior at 240°C and 5V/μm was initially dominated by tunneling current through grain boundaries and later by the decreased resistance at the ceramic/internal interface induced by accumulation of oxygen vacancy near the cathode.

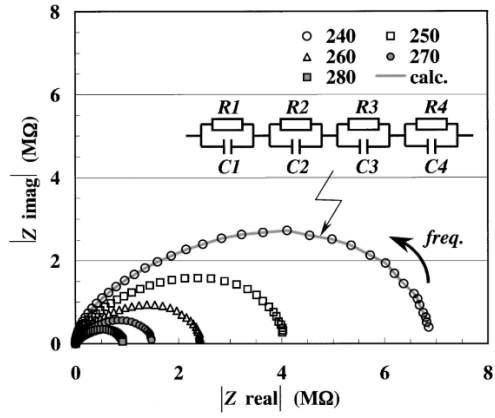

Figure 20. Complex impedance data of a X7R MLCC measured at various temperatures as shown in the legend. The solid line shows the fitted impedance for a 4-RC electrical equivalent circuit. Reproduced with permission from ref. [180].

To ascertain the electrical characteristics of a material in its degraded state, the sample can be degraded *in situ* while concurrently mapping the conductivity distribution. This can be achieved by using a DC voltage to induce degradation, while alternating current (AC) frequency sweeps map the conductivity distribution. The evaluation of local conductivity fundamentally relies on the relationship between the conductivity ($\sigma$), the relative permittivity ($\varepsilon_r$), the permittivity of free space ($\varepsilon_0$), and the relaxation frequency ($f_r$) [133,439,440]:

$$f_r = \frac{\sigma}{2\pi\varepsilon_0\varepsilon_r} \qquad (29)$$

This correlation is pertinent for a parallel arrangement of a capacitor and a resistor, which is typified by a Debye peak in the imaginary component of the electric modulus M"(f) at the relaxation frequency $f_r$. When the specimen has two regions with distinct conductivities connected in series, the modulus profile displays two Debye peaks in the frequency domain [441–444]. The degradation kinetics of undoped and Fe-doped STO are depicted in Figure 21, which shows the real component of admittance (conductance), Y', as a function of time, along with M"(f) spectra taken during the degradation process. The time dependent admittance plot, Y', demonstrates an escalation in the conductivity of the STO, culminating in a stable degraded state. Concurrently, the M"(f) spectra changes at various stages of degradation. Specifically, the degradation in Fe-doped STO is marked by the emergence of two distinct conductive regions at the anode and cathode, giving rise to separate M"(f) peaks. The magnitude of these peaks is indicative of the size of the respective conductive area. The minor peak at lower frequencies is associated with the less conductive anode region, while the larger peak corresponds to the more conductive cathode region. The change in conductance over time for each area can be individually monitored by selecting appropriate frequencies. Conversely, undoped STO does not split into two peaks during degradation. Only a slight reduction in peak amplitude is noted. A comparison of the degradation in undoped STO samples with different dielectric thicknesses, all exposed to equivalent electrical fields, shows that samples with greater thickness exhibit a more pronounced increase in conductivity as a result of field stress. Intriguingly, the degradation timeline is largely unaffected but the dielectric thickness, in contrast to acceptor-doped STO [143].

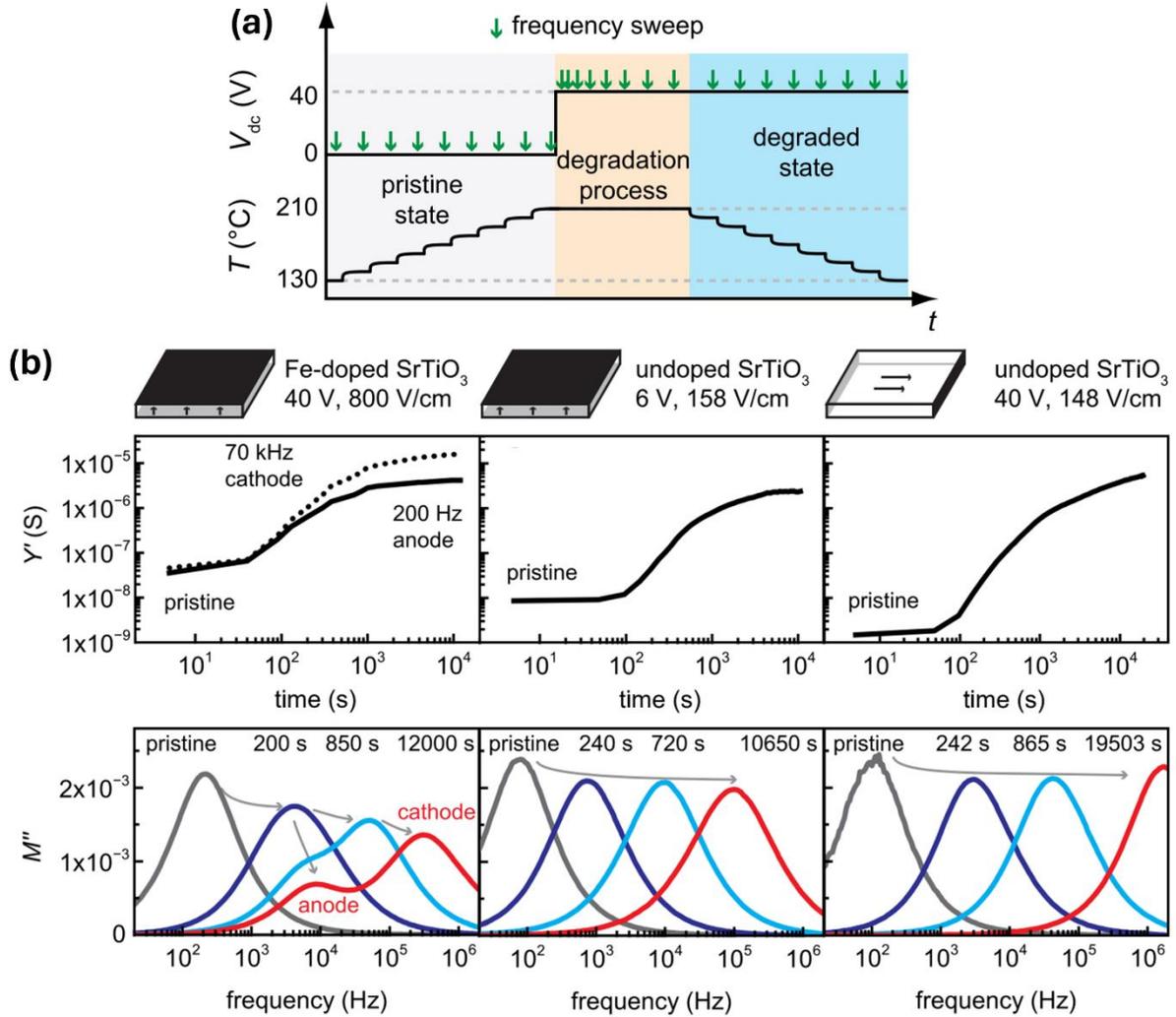

Figure 21. (a) A time series of DC voltage, frequency sweeps, and temperature variations were employed to examine the initial pristine state, the degradation process, and the degraded samples. (b) The degradation of both Fe-doped and undoped STO samples in the planar geometry, as well as undoped STO in the lateral geometry, is depicted. The degradation is visualized through admittance as a function of time at a constant frequency and as the imaginary part of the modulus M″ as a function of frequency at specific time intervals. Reproduced with permission from ref. [143].

### 4.3.2 Thermally Stimulated Depolarization Current (TSDC)

Thermally Stimulated Depolarization Current (TSDC) is a powerful technique used for investigating the relaxation kinetics of polarizable defects, as well as for detecting changes in defect dipole complexes, electron trapping, and the development of ionic space charge, both intergranular and transgranular [132,249,442,445–456]. Foundational theoretical interpretation was provided by Bucci et al. [457]. Figure 22.a presents a schematic representation of the standard TSDC experiment. The sample undergoes polarization at a temperature ($T_p$) under a constant electric field ($E_p$) for a predetermined poling duration, which may range from minutes to hours depending on the nature of the sample. Subsequently, the sample undergoes rapid cooling to a lower temperature ($T_0$), effectively preserving the polarized defects. The dielectric is short-circuited at $T_0$, and then the sample is subjected to gradual heating at a constant heating rate. This controlled heating process prompts the depolarization of each metastable state, resulting in the generation of a depolarization current, which can be measured using a pA meter [458].

Together EPR and TSDC techniques are pivotal for identifying and analyzing defect complexes in materials. EPR is particularly effective when an oxygen vacancy is adjacent to an acceptor ion on a nearest-neighbor site. However, if the vacancy shifts to the next nearest-neighbor site, the EPR signal vanishes. In systems with low concentrations of acceptor defect centers, the likelihood of oxygen vacancies being nearest neighbors to acceptor ions is reduced, as they are outnumbered by available sites surrounded only by $Ti^{4+}$ ions. Without a specific mechanism promoting the formation of defect complexes, oxygen vacancies may occupy random lattice sites, reducing the fraction of defect complexes detectable by EPR to the total number of acceptor defects not forming associates [343,370]. In contrast, TSDC offers insights into the physical origins of each observed peak through the study of the electric field's influence on the peak's maximum temperature ($T_{max}$), while the material is heated at a constant rate.

As the heating rate increases, the depolarization peaks narrow and shift to higher temperatures [459]. $T_{max}$ decreases with poling field when associated with trapped charges, increases for space charge, and remains constant when the TSDC current arises from defect dipoles [447,460]. $T_{max}$ depends on characteristic relaxation time for polarized charges and is given by [450]:

$$T_{max} = \left[\frac{E_a}{k_B}\beta\tau_0 exp\left(-\frac{E_a}{k_B T_{max}}\right)sinh\left(\frac{qaE_p}{2k_B T_p}\right)\right] \qquad (30)$$

Moreover, integration of depolarization current peaks and subsequent curve fitting yield insights into relaxational kinetics, including the concentration of mobile charge carriers or dipoles and their respective activation energies. An equation adapted from thermoluminescent glow curve theory is often employed to fit TSDC peaks originating from single relaxation of trapped charges:

$$I(T) = n_o s \, exp\left(-\frac{E_a}{k_B T}\right) exp\left[-\frac{s}{\beta}\int_{T_o}^{T}\left(-\frac{E_a}{k_B T'}\right)dT'\right] \qquad (31)$$

$$s = \beta\frac{E_a}{k_B T_{max}^2 exp\left(-\frac{E_a}{k_B T_{max}}\right)} \qquad (32)$$

where $T_{max}$, and $\beta$ are the temperature corresponding to the peak maximum and the heating rate, respectively, while $n_o$ is dipole concentration and $s$ is a geometric factor that depends on the dipole orientation [428]. Moreover, the total charge ($Q_{TSDC}$) released as a result of the displacement of mobile ions is determined by [442]:

$$Q_{TSDC} = 2a\vartheta Nqt_p exp\left(-\frac{H}{k_B T}\right) sinh\left(\frac{qaE_p}{2k_B T_p}\right) \tag{33}$$

where $t_p$ and $H$ are the time and depth of potential wells, respectively.

Three methods are commonly employed to determine the activation energy of a TSDC peak: the rise of initial current, the Full Width Half Maximum (FWHM) of the TSDC peak, and the heating rate dependence of the TSDC peak position [249]. In the first method, the temperature rise of the initial current is controlled by the first exponential term of equation (31). Plotting ln $J_D$ versus 1/T results in a linear relationship, allowing for the calculation of the activation energy from the slope of this line. The second approach involves using the FWHM of the TSDC peak to determine the activation energy, given by the equation:

$$E_a = 2.3 k_B \left(\frac{T_{max}^2}{\Delta T_{1/2}}\right) \tag{34}$$

where $\Delta T_{1/2}$ is the FWHM of the TSDC peak. The third, and often more accurate, method involves describing the temperature dependence of the relaxation time using the Arrhenius equation, allowing extraction of the activation energy of the TSDC peak [455]:

$$\ln\left(\frac{T_{max}^2}{\beta}\right) = \frac{E_a}{k_B T_{max}} + \ln\left(\frac{\tau_0 E_a}{k_B}\right) \tag{35}$$

where $E_a$ and $\tau_0$ are the activation energy of the barrier height controlling diffusion and the relaxation time, respectively. The activation energies can be calculated from the slopes of a $\ln\left(\frac{T_{max}^2}{\beta}\right)$ vs. $\frac{1}{T_{max}}$ plot.

In TSDC analysis, resolving overlapping relaxation processes presents a significant challenge. Peak cleaning is a well-established approach, adapted from thermoluminescence experiments, to tackle this issue. This method involves conducting a thermal cycle that initially records the entire discharge curve. In a subsequent cycle, the material is heated to a specific temperature that effectively discharges the lower temperature peak, known as peak 1, while leaving the higher temperature 2, unaffected. After this, the sample is cooled down and then reheated, allowing for the isolated discharge of peak 2. This results in a clearer and more distinct discharge curve for peak 2 [358], as shown in Figure 22.c. Complementing this is a method developed by Bucci et al. [373], tailored for TSDC measurements. This technique relies on selectively polarizing the material at a temperature where one type of dipole reaches saturation while the other remains in a random state. By carefully choosing the temperatures and polarization intervals, it is possible to saturate each type of dipole separately and measure them in isolation. However, while both effectively separate overlapping peaks, they also risk truncating the original distributions. This alteration results in TSDC peaks that reflect the measurement conditions employed rather than the inherent characteristics of the material [358]. Figure 22.d [249] shows use of TSDC to probe the nature and concentration of defects that lead to electrical degradation in PZT films. TSDC measurements for Mn-doped PZT films exhibit a singular depolarization peak, with a shift of $T_{max}$ towards higher temperatures in response to increasing poling electric fields. This suggests that the TSDC peak arises from space-charge polarization, which can be attributed to the migration of oxygen vacancies in the films. The activation energy associated with the TSDC peak was evaluated using the initial rise, Full Width at Half Maximum (FWHM), and heating rate methods. The activation energies determined via three separate methods span from 0.58 to 0.73 eV. The estimation derived from the heating rate method is posited to be more precise as the dependence of the peak temperature on the heating rate is exclusively related to the relaxation kinetics of trapped charges, defect dipoles, or ionic spaces, rather than the background current produced by the pyroelectric effect. The activation energy values obtained from these methods concur with the findings for oxygen vacancy migration reported by Wang et al., which is approximately 0.58 eV[249,256].

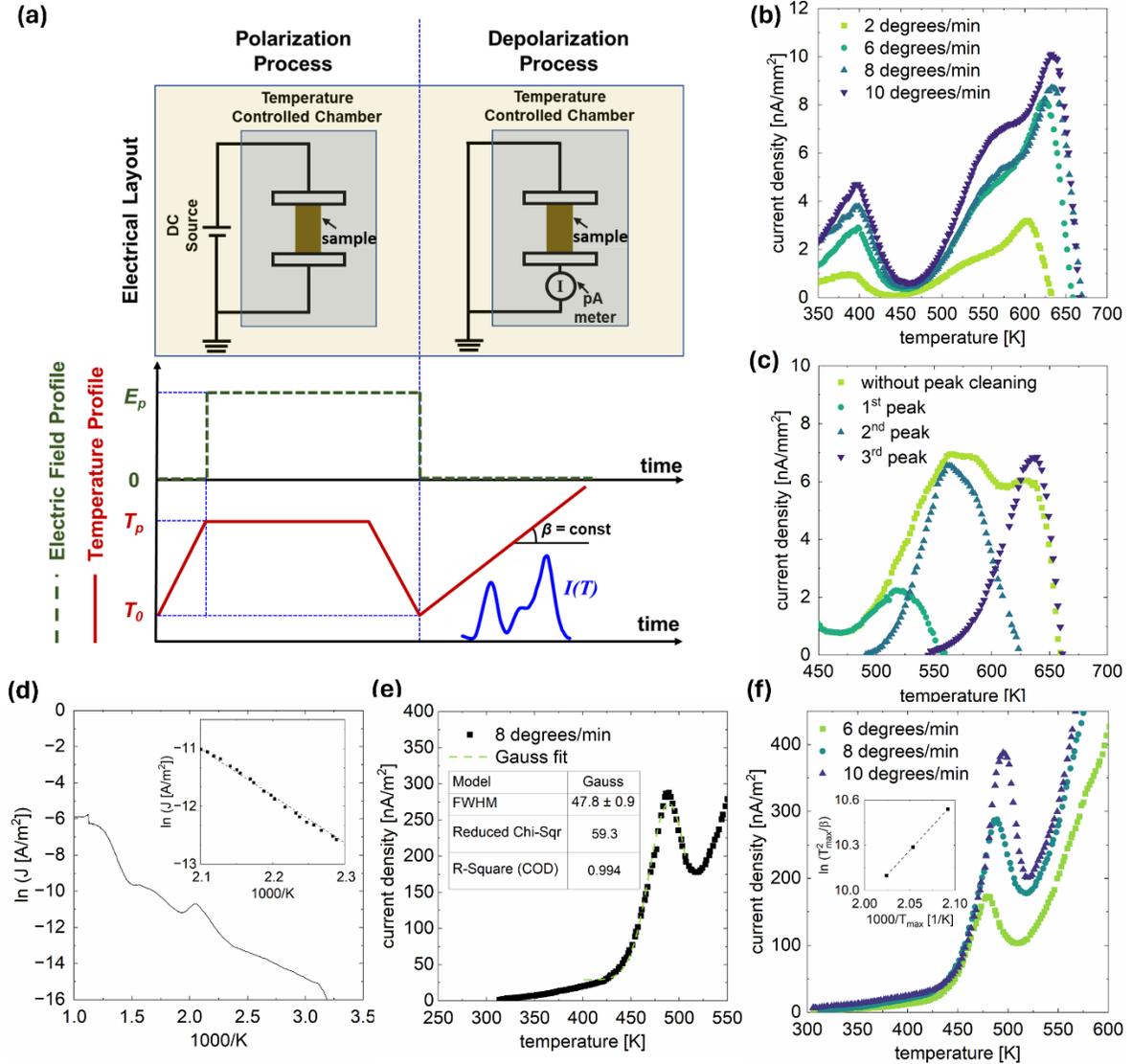

Figure 22. (a) Schematic of typical test conditions for a TSDC experiment (reproduced with permission from ref. [458]), (b) evolution of TSDC peaks with increasing heating rate, (c) peak cleaning method data, and estimation of activation energies for TSDC peaks arising from 2 mol% Mn doped PZT films using: (d) the initial rise model, (e) the FWHM method, and (f) different heating rates (6-10°C/min). The films were degraded under an electric field of 275 kV/cm at 180°C for 12 hrs (redrawn from data in ref. [250]).

TSDC can also be used to understand the association and dissociation of oxygen vacancy -acceptor complexes under an applied electric field. The co-doping of RE dopants and acceptors in BTO ceramics can enhance the reliability of the dielectric, and one of the mechanisms is related to the association of the oxygen vacancies, thereby limiting the total mobile vacancy concentration. A theoretical study of these types of complexes was conducted by Honda et al. [123]; they show that different configurations, e.g. nearest neighbour and next nearest neighbour configurations, have different energies. Associated defects can orient relative to the field and therefore behave as a defect dipole with respect to the TSDC. Optimized polarization fields, temperatures, and times can maximize the alignment of the defect dipoles. Under more aggressive polarization conditions, the defect dipole peak often drops in intensity, coupled with a corresponding increase in a space charge peak associated with in grain ionic space charge at the grain boundaries. Figure 23.a shows a schematic of the dissociation and changes in the TSDC, Figure 23.b shows data for the TSDC peaks, and the kinetic studies for different temperatures and electric fields for a co-doped BTO system [51]. Kinetic models of the data allow the activation energy for dissociation of the oxygen vacancy from the defect complex to be determined. Assuming that dissociation is a pseudo-first order reaction, the kinetics of the dissociation is given by the following equation:

$$\ln\left(\frac{J}{J_{max}}\right) = -\mathrm{K}t \tag{36}$$

where $J_{max}$ is the saturation of the defect depolarization current, K is the temperature-dependent rate constant of the first-order reaction.

Inoue et al. studied the dissociation kinetics in BTO co-doped with RE (Gd, Y, Dy, Ho) and Mg. These RE dopants are amphoteric, such that they can occupy either A- or B-sites in the lattice, while the Mg prefers B-site occupancy based on ionic size [3]. This preference results in different interaction energies depending on the RE site occupancy. TSDC results revealed that after annealing the degraded sample at 600°C, the defect dipole depolarization peaks shifted to lower temperatures. It was observed that the degradation process reduced the dissociation energy by 0.2 eV compared to the initial state associated with random defect dipoles formed by cooling from sintering conditions, prior to electrical degradation. This change was not recovered even after the 600°C annealing. It is believed that the defect dipoles are typically in the nearest neighbor configuration in as-prepared ceramics. The reduction in dissociation energy post-degradation suggests that the oxygen vacancies do not revert to the most stable energy state for association. Instead, they likely occupy next nearest neighbor positions. This results in lower energies for dissociation compared to the initial state. Figure 23.c represents TSDC spectra before and after degradation and annealing, and the corresponding change in the activation energy. Each TSDC measurements required re-establishing the defect association to its lowest energy state, achieved by annealing above 1000°C and cooling at consistent rates.

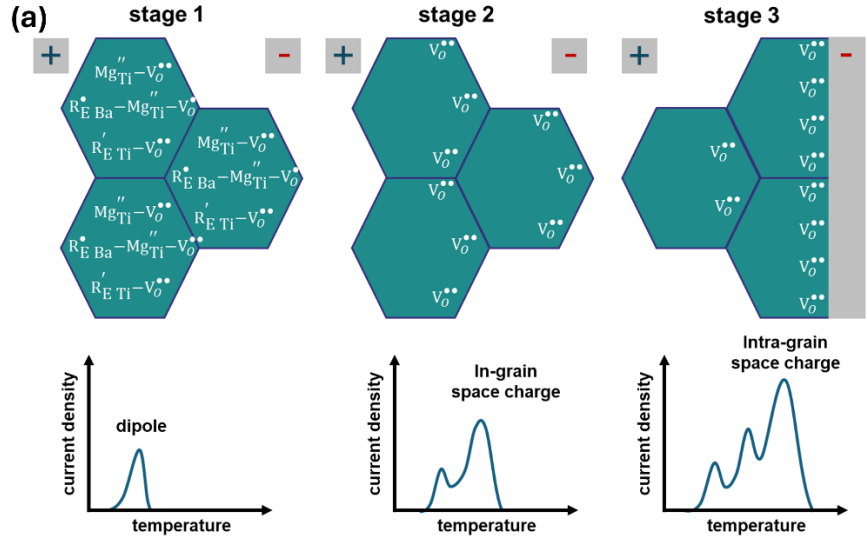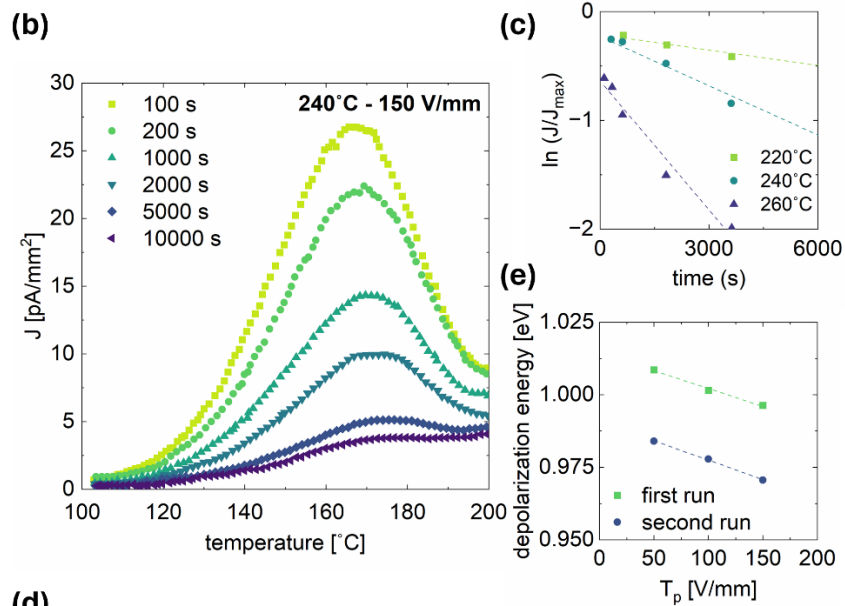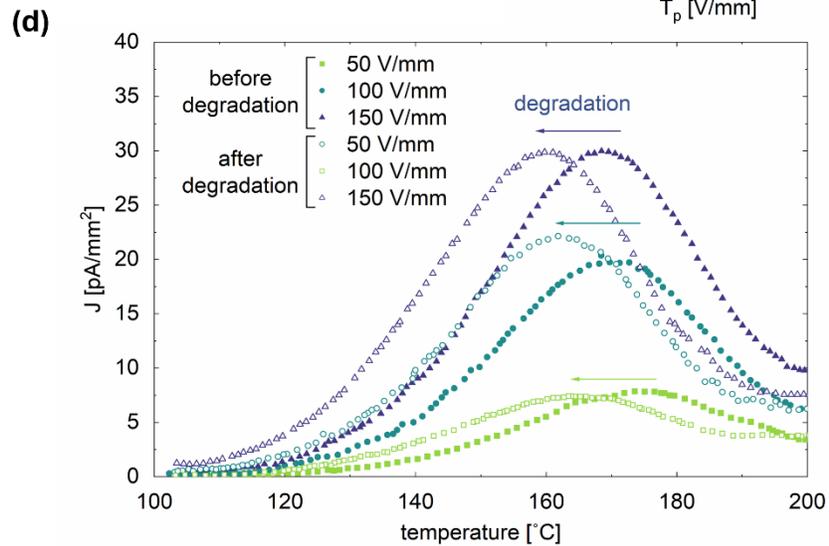

Figure 23. (a) Schematic of dissociation of defects during electrical degradation and corresponding TSDC spectra, (b) TSDC spectra for $(Ba_{0.94}Dy_{0.06})(Ti_{0.97}Mg_{0.03})O_3$ with $E_p$ = 150 V/mm, $T_p$ = 240°C, and $t_p$ = 100–10000 s, (c) polarization time dependence of $\ln(J/J_{max})$ with $E_p$ = 100 V/mm, $T_p$ = 220-260°C, and $t_p$ = 100–10800 s (redrawn from data in ref. [51]), (d) TSDC spectra before and after degradation and (e) corresponding dissociation energy plot.

TSDC analysis of ferroelectrics encounters challenges arising from background currents caused by the pyroelectric effect. Wu and Sayer [461] delved into the aging association with the TSDC peak linked to oxygen vacancies in piezoelectric thin films by implementing a two-probe method within a specialized experimental setup to distinguish TSDC signals from pyroelectric effects.

Liu et al. explored the evolution of trap depth and trap density during degradation in single and polycrystal material Fe-doped STO using TSDC [132,442]. In acceptor-doped BTO, Yoon et al. [163,445,446,454] found two space charge peaks associated with oxygen vacancy relaxation: one with a lower activation energy for within-grain diffusion and another with higher activation energy for across grain boundaries diffusion. Yousefian et al. [458,462] demonstrated that TSDC may serve as a rapid and efficient quality control technique for assessing the quality of MLCCs.

### 4.3.3 Highly Accelerated lifetime Test (HALT)

The initial usage of the Highly Accelerated Lifetime Test (HALT) technique marked a milestone in the field of reliability engineering. HALT accelerates the degradation process by subjecting the component to extreme stress conditions, far beyond its standard operational parameters. During this testing, the leakage current is monitored; typically, it increases gradually until dielectric breakdown is observed. Extrapolation of the lifetime to standard operating conditions offers a perspective on the component's long-term performance and reliability. HALT can help identify various degradation and breakdown mechanisms, effectively distinguishing between issues caused by manufacturing processes (extrinsic breakdown) and those stemming from material defects (intrinsic breakdown) [168,169,463–475].

The foundational empirical model for HALT, known as the P-V or Eyring model, was proposed by Prokopowicz and Vaskas [476], and can be expressed as:

$$\frac{t_1}{t_2} = \left(\frac{V_2}{V_1}\right)^n exp\left[\frac{E_a}{k_B}\left(\frac{1}{T_1} - \frac{1}{T_2}\right)\right] \tag{37}$$

where $t_1$ and $t_2$ are the MTTFs measured at voltages $V_1$ and $V_2$ and at corresponding temperatures $T_1$ and $T_2$, respectively. The parameters $n$ and $E_a$ represent the electric-field acceleration factor, and activation energy governing the degradation process, respectively. While this model has been widely used and adapted, it is important to note that the factor $n$, is not a physical constant and may vary under different test conditions [465,477–479]. This variability has been a point of contention, as it can significantly affect the model's accuracy [3,458,480,481]. In PME MLCCs, *n* values are often treated as constants. Conversely, in BME MLCCs, the *n* values are not constant, posing challenges in lifetime prediction and necessitating understanding of the non-linear behavior of *n*.

To overcome the limitations of the Eyring model, Randall et al. [134] proposed the tipping point lifetime model. This physical model is based on the critical space charge density accumulation at the cathode interface and factors in local fields [134]. In the tipping point model, the lifetime of MLCCs is:

$$t_{crit} = \frac{\rho_{crit}}{a\vartheta N q}\left[\exp\left(-\frac{E_a}{k_B T}\right)\sinh\left(\frac{qaE_{app}}{2k_B T}\right)\right]^{-1} \tag{38}$$

where $t_{crit}$ and $\rho_{crit}$ represent the predicted lifetime and the critical space charge density at the cathode interface, respectively. *a, ϑ, N, q, T, $E_a$,* and *$E_{app}$* are the characteristic hopping distance, characteristic jump frequency of the oxygen vacancy, oxygen vacancy concentration, oxygen vacancy charge, polarization temperature, the activation energy of the barrier height controlling diffusion, and applied electric field, respectively [134,482]. Morita et al. [480] adapted the tipping point model to improve its low field predictions, incorporating time-dependent internal depolarization field development. The tipping point model has been shown to be more systematic and predictable across various testing conditions than the Eyring model. Nevertheless, it's important to note that both models necessitate the collection of extensive experimental data to be effectively applied and validated [458,482]. To overcome the limitations of existing models, Yousefian et al. [482] developed a physically-based machine learning model to predict the MTTF of MLCCs. This model utilizes a transfer learning framework that incorporates the principles underlying existing models, thereby enhancing its accuracy and ensuring stable performance under various testing conditions, regardless of the failure mechanisms at play, even when working with a limited dataset.

Both industry and academia frequently rely on MTTF values for estimating the lifetime of electronic components like MLCCs under operating conditions [134,169,465,467,481,483–486]. However, this approach has its limitations, as MTTF does not cover the full distribution of failure times. This can lead to unexpected failures. To achieve more precise lifetime predictions, it is crucial to consider a range of influencing factors such as voltage, temperature, and humidity, and apply these across the entire failure time spectrum. Incorporating this holistic view of failure time distribution into the analysis offers a more comprehensive understanding. Furthermore, the bathtub curve commonly employed for hazard rate curves in electronic components might not reflect reality [487–490]. Some researchers have proposed a roller-coaster-like shape for these curves, indicative of latent failures and multiple failure mechanisms [491,492]. Recognizing the complexities involved is crucial for the reliability analysis of electronic components such as MLCCs. This understanding leads to the development of more precise models and a deeper insight into the underlying failure mechanisms.

HALT, in conjunction with the Eyring model, has been extensively employed across various materials systems to uncover degradation mechanisms and determine corresponding activation energies [236,238,265,493,494]. This methodology has been instrumental in the study of PZT films. The lifetime of the PZT films rises with higher donor doping levels, which corresponds to a reduction in mobile oxygen vacancy concentration. The calculated activation energies derived from HALT measurements closely align with the effective barrier height values obtained from leakage current measurements, implying a correlation between lifetime and Schottky emission phenomena [250]. This points to the critical role of interfacial defect chemistry in governing the electrical degradation behavior of these films. Consequently, HALT has emerged as a vital tool in the development of more resilient and reliable PiezoMEMS devices, designed to operate effectively under extreme electric fields and temperatures, bolstering their suitability for a wide range of applications.

## 5   Summary and Perspectives

Perovskite-based dielectrics and piezoelectrics, including STO, BTO, BST, and PZT, are anticipated to increase in market share over the foreseeable future as outlined in several market forecasts for MLCCs, actuators, sensors, transducers, PTCR, ferroelectric memory, and optoelectronic devices. The development of perovskite materials that maintain robust dielectric and piezoelectric responses under high electric fields and temperatures is essential for their application in electronic circuits. With market expansion and the advent of novel applications, the number of suppliers increases, placing pressure on the supply chain and potentially impacting the quality of products due to the need for rapid and large-scale production. The trend toward more compact electronic devices also necessitates miniaturization of circuits, which could raise concerns over electrical reliability. In high-temperature power electronics, the use of capacitive elements introduces the challenge of increased dielectric losses, affecting circuit efficiency and overall energy consumption. Moreover, electrical degradation leading to dielectric breakdown can result in failures at the system or component level. Thus, maintaining the integrity of capacitive components is crucial, requiring an understanding of defect chemistry, charge transport mechanisms, and electrical reliability. This paper describes these charge transport mechanisms and their role in leakage current and electrical degradation. The importance of structural, optical, thermal, and electrical characterization is emphasized to elucidate the relationship between defect chemistry, charge carrier dynamics, and the electrical reliability of perovskite ceramics.

In perovskite ceramics, which act as mixed electronic-ionic conductors, electrical degradation is primarily caused by the electromigration of oxygen vacancies moving from the anode toward the cathode. The electrical degradation can be mitigated by optimizing sintering conditions, engineering grain boundaries/interfaces, controlling oxygen partial pressures, and implementing specific doping strategies. Increased defect concentration at grain boundaries, coupled with acceptor ion segregation during sintering, forms double Schottky barriers that enhance space charge depletion, further inhibiting oxygen vacancy migration during electrical degradation. Likewise, in core-shell MLCCs, acceptor ion segregation within the shell layer during the sintering process generates a space charge layer. This layer serves as an additional barrier to the oxygen vacancy migration, thereby increased the lifetime. The optimization of the internal electrode interface through controlling annealing atmosphere is pivotal in moderating electron injection at the cathode within MLCCs.

In thin film perovskites, electrical degradation is strongly influenced by defect chemistry at the electrode/film interface. A higher concentration of oxygen vacancies is typically found near these interfaces, attributed to their lower formation energies at surfaces. In particular, PbO loss in PZT films during annealing significantly impacts the concentration of oxygen vacancies near the surface, as well as their distribution throughout the film's thickness. To mitigate electrical degradation in perovskite films, it is crucial to minimize the effective charge density resulting from the accumulation of oxygen vacancies near the cathode interface. This can be accomplished by: (1) Conducting post-annealing in a controlled PbO atmosphere to reduce the oxygen vacancy concentration within the PZT film. (2) Modifying the interfacial defect chemistry by inserting a layer of acceptor-doped film between the cathode and the bulk film.

The development of new components and compositions necessitates rigorous quality controls that are in line with industry standards, ensuring accurate communication of batch performance statistics. The techniques and strategies reviewed in this paper are useful towards that goal. Detailed attention is being given to understanding interfaces, grain sizes, compositional development, and composition distributions within the microstructures. Challenges can come in scaling to mass production, chemical mixing, temperature and atmosphere gradients, and flow rates within reactors can produce subtle but important variations that broaden-the population distribution for properties such the MTTF. This will not only perturb the MTTF but also the variance of the population. Currently, there are models that accurately extrapolate–mean lifetimes under use conditions from accelerated lifetime MTTF measurements. However, the extrapolation of the variance is a challenge. For very high voltages or temperatures, components can be overstressed, and induce some failure mechanisms that are not relevant to the operational conditions of components. However, at lower stress conditions within a single mode failure, prediction of the variance in HALT to operational conditions is unclear.

The variance is critical, as weak components within the population will start to undergo degradation during use, with commensurate increases in Joule heating. In densely packed circuits, any local increase in temperature will influence the nearby circuit components, triggering increased failure rates that depend on the circuit layout. With the complexity of circuits increasing, the value of understanding the science of failure at the individual component level and at the packaged circuit level will continue to increase.

## 6    Acknowledgements

This work is supported by the National Science Foundation, as part of the Center for Dielectrics and Piezoelectrics (CDP) under Grant Nos. IIP-1841453 and IIP-1841466".

functional point defect in SrTiO$_3$. Phys Rev Mater 2018;2:1–7. https://doi.org/10.1103/PhysRevMaterials.2.060403.

[389]   Chen L, Fu Q, Jiang Z, Xing J, Gu Y, Zhang F, et al. Cathodoluminescence evaluation of the degradation of Mg, Ca and Dy Co-doped BaTiO$_3$ Ceramics. J Eur Ceram Soc 2021;41:7654–61. https://doi.org/10.1016/j.jeurceramsoc.2021.08.017.

[390]   Rohj RK, Hossain A, Mahadevan P, Sarma DD. Band gap reduction in ferroelectric BaTiO$_3$ through heterovalent Cu-Te co-doping for visible-light photocatalysis. Front Chem 2021;9:1–8. https://doi.org/10.3389/fchem.2021.682979.

[391]   Gatasheh MK, Daoud MS, Kassim H. Bandgap narrowing of BaTiO$_3$-based ferroelectric oxides through cobalt doping for photovoltaic applications. Materials 2023;16:7528. https://doi.org/10.3390/ma16247528.

[392]   Olatunji SO, Owolabi TO. Barium titanate semiconductor band gap characterization through gravitationally optimized support vector regression and extreme learning machine computational Methods. Math Probl Eng 2021;2021. https://doi.org/10.1155/2021/9978384.

[393]   Choi M, Oba F, Tanaka I. Electronic and structural properties of the oxygen vacancy in BaTiO$_3$. Appl Phys Lett 2011;98. https://doi.org/10.1063/1.3583460.

[394]   Gao H, Sahu S, Randall CA, Brillson LJ. Direct, spatially resolved observation of defect states with electromigration and degradation of single crystal SrTiO$_3$. J Appl Phys 2020;127. https://doi.org/10.1063/1.5130892.

[395]   Zhang J, Walsh S, Brooks C, Schlom DG, Brillson LJ. Depth-resolved cathodoluminescence spectroscopy study of defects in SrTiO$_3$. Journal of Vacuum Science & Technology B: Microelectronics and Nanometer Structures Processing, Measurement, and Phenomena 2008;26:1466–71. https://doi.org/10.1116/1.2918315.

[396]   Ihrig H, Klerk M. Visualization of the grain-boundary potential barriers of PTC-type BaTiO$_3$ ceramics by cathodoluminescence in an electron-probe microanalyzer. Appl Phys Lett 1979;35:307–9. https://doi.org/10.1063/1.91119.

[397]   Ihrig H, Hengst JHT, Klerk M. Conductivity-dependent cathodoluminescence in BaTiO$_3$, SrTiO$_3$ and TiO$_2$. Zeitschrift Für Physik B Condensed Matter 1981;40:301–6. https://doi.org/10.1007/BF01292846.

[398]   Yacobi BG, Holt DB. Cathodoluminescence Microscopy of Inorganic Solids.